\documentclass{imsart}
\usepackage{fullpage}

\RequirePackage{amsthm,amsmath,amsfonts,amssymb}
\RequirePackage[authoryear]{natbib}
\RequirePackage[colorlinks,citecolor=blue,urlcolor=blue]{hyperref}
\RequirePackage{graphicx}
\usepackage{algorithm2e}[ruled]

\startlocaldefs
\theoremstyle{remark}
\newtheorem{model}{Model}

\newcommand{\crps}{\operatorname{crps}}
\newcommand{\crpss}{\operatorname{crpss}}
\newcommand{\ts}{\operatorname{test}}
\newcommand{\tr}{\operatorname{train}}
\newcommand{\lpd}{\operatorname{lpd}}

\newcommand{\unif}{\operatorname{Uniform}}
\newcommand{\pois}{\operatorname{Poisson}}
\newcommand{\nb}{\operatorname{Neg-Bin}}
\newcommand{\gam}{\operatorname{Gamma}}
\newcommand{\norm}{\operatorname{Normal}}
\newcommand{\e}{\operatorname{Exp}}
\newcommand{\te}{\operatorname{Trunc-Exp}}
\newcommand{\var}{\operatorname{Var}}
\newcommand{\rep}{\operatorname{trans}}

\endlocaldefs

\begin{document}

\begin{frontmatter}
\title{Modelling superspreading dynamics and circadian rhythms in online discussion boards using Hawkes processes.
}

\runtitle{Modelling online discussions} 


\begin{aug}
\author[A]{\fnms{Joe}~\snm{Meagher} \ead[label=e1]{joe.meagher@hyman.co.uk
} \orcid{0000-0002-2650-2506}}
\and
\author[A]{\fnms{Nial}~\snm{Friel}\ead[label=e2]{nial.friel@ucd.ie} \orcid{0000-0003-4778-0254}}
\address[A]{Insight Centre for Data Analytics, School of Mathematics and Statistics, University College Dublin. \\
    \printead[presep={}]{e1} \printead[presep={,\ }]{e2}}

\end{aug}

\begin{abstract}
Online boards offer a platform for sharing and discussing content, where discussion emerges as a cascade of comments in response to a post. Branching point process models offer a practical approach to modelling these cascades; however, existing models do not account for apparent features of empirical data. We address this gap by illustrating the flexibility of Hawkes processes to model data arising from this context as well as outlining the computational tools needed to service this class of models. 
For example, the distribution of replies within discussions tends to have a heavy tail.
As such, a small number of posts and comments may generate many replies, while most generate few or none, similar to `superspreading' in epidemics.
Here, we propose a novel model for online discussion, motivated by a dataset arising from discussions on the \texttt{r/ireland} subreddit, that accommodates such phenomena and develop a framework for Bayesian inference that considers in- and out-of-sample tests for goodness-of-fit.
This analysis shows that discussions within this community follow a circadian rhythm and are subject to moderate superspreading dynamics.
For example, we estimate that the expected discussion size is approximately four for initial posts between 04:00 and 12:00 but approximately 2.5 from 15:00 to 02:00. We also estimate that 58--62\% of posts fail to generate any discussion, with 95\% posterior probability.
Thus, we demonstrate that our framework offers a general approach to modelling discussion on online boards.
\end{abstract}


\end{frontmatter}


\section{Introduction}
\label{sec:intro}

Online boards provide a forum for sharing and discussing content, a process crucial to forming, organising, and preserving online communities.
Discussion on these boards forms cascade-like data structures, where one event triggers subsequent events.
Such data are ubiquitous online and have proven a popular research topic in Computational Social Science (see \cite{zhou2021survey} for a review).
In this context, modelling the trajectory of individual cascades (i.e. understanding and predicting how content popularity changes over time) remains a fundamental and challenging question \citep{watts2002simple, cheng2014can, moniz2019review, zhou2021survey, bollenbacher2021challenges}.

Human behaviour is often characterised by short bursts of activity followed by long periods of inactivity \citep{barabasi2005origin}, and such irregular dynamics complicate the structure of cascades generated by social interaction.
That said, generative models based on self-exciting point processes, where the occurrence of past events makes future events more likely (see, e.g. \cite{daley2003introduction}), have proven effective.
In particular, Hawkes processes \citep{hawkes1971point, hawkes1971spectra, hawkes1974cluster} underpin models for retweet cascades on Twitter \citep{zhao2015seismic, kobayashi2016tideh, rizoiu2018sir, kong2020describing} and discussion cascades on Reddit \citep{medvedev2019modelling, krohn2019modelling}.
Such models have several advantages over competing methods.
They are straightforward to simulate \citep{moller2006approximate} and perform well on prediction tasks \citep{kobayashi2016tideh, krohn2019modelling}.
Furthermore, they are amenable to rigorous analysis \citep{o2020quantifying} and methods for likelihood-based inference are well established \citep{mohler2011self,rasmussen2013bayesian}.

Despite the appeal of these branching process models, existing implementations are limited in the following sense.
In general, analyses have focused on developing methods for \textit{ex-post} prediction (forecasting the trajectory of the cascade at any time after it is seeded) and analysing cascades that have already reached some popularity threshold \citep{zhao2015seismic, kobayashi2016tideh, medvedev2019modelling}.
Such cascades are relatively rare occurrences, and so these methods do not offer a principled approach to studying cascades more generally \citep{watts2002simple, goel2012structure}.

A second issue is that these branching processes typically assume that offspring are Poisson distributed \citep{kobayashi2016tideh, rizoiu2018sir, medvedev2019modelling, krohn2019modelling}. However, \cite{medvedev2019modelling} demonstrated that this assumption does not hold for discussions on Reddit --- some events give rise to many offspring while the majority generate few or none. 
This phenomenon is analogous to superspreading in epidemics, which emerges from the heterogeneous transmission of infectious disease \citep{lloyd2005superspreading, meagher2022assessing}.
That is to say, when `infectiousness' varies from one individual to the next, the distribution of offspring becomes over-dispersed.
Accounting for this behaviour may provide models that more accurately describe cascade dynamics.

This paper addresses each of these shortcomings in an analysis of discussions on the \texttt{r/ireland} subreddit.
To this end, we propose a novel generative model for online discussions which includes a circadian rhythm, allows for over-dispersed offspring distributions and admits simpler Hawkes process models as special cases.
By adopting a model for the spread of infectious disease in this new context, our generative model describes online discussions which follow a `bursty' trajectory \citep{lloyd2005superspreading}. 
As such, we develop a hierarchical model for online discussion where all discussions are considered jointly within a single modelling framework, and Monte Carlo methods allow for efficient Bayesian parameter inference \citep{carpenter2017stan}, evidence estimation \citep{gelman1998simulating, gronau2017tutorial}, and model assessment \citep{gneiting2007strictly, vehtari2017practical}. 
Thus, our methodology is suitable for explanatory analysis as well as both \textit{ex-ante} (up to and including the moment that the discussion is seeded) and \textit{ex-post} prediction tasks.

There are several advantages to this approach.
Firstly, our analysis provides new insight into the \texttt{r/ireland} community. 
For example, we estimate that content submitted between 04:00 and 12:00 has a posterior mean discussion size of approximately 4, while that submitted between 15:00 and 02:00 has a posterior mean size of approximately 2.5.
In addition, we estimate that 58--62\% of posts generate no further discussion, with 95\% posterior probability.
More generally, our framework applies to discussions on any forum and sits squarely within the integrated modelling approach to Computational Social Science outlined by \cite{hofman2021integrating}, where analyses combine aspects of both explanatory and predictive modelling to generate new insight. 
As such, our methods offer an explanation for sampled data, which we validate with in- and out-of-sample testing.
Finally, this work highlights Hawkes process models' flexibility and broad applicability.

The remainder of our paper is laid out as follows.
In Section \ref{sec:background}, we introduce our data set and perform exploratory analysis before presenting the necessary background on marked point processes and epidemic models.
We present the generative model in Section \ref{sec:generative_model} along with our approach to parameter inference in \ref{sec:parameter_inference} and model assessment in \ref{sec:model_assessment}.
Section \ref{sec:results} presents the results of our analysis of discussions on the \texttt{r/ireland} subreddit, and we conclude with a discussion in Section \ref{sec:discussion}.

\section{Background}
\label{sec:background}

\subsection{Sharing and discussing content on Reddit}
\label{sec:reddit}

Reddit, the sixth most-visited website on the internet \citep{semrush}, is an archetypal online board. 
As described by \cite{medvedev2017anatomy}, the website divides into \textit{subreddits} created by users to host discourse on a particular topic. 
Users are responsible for both generating and moderating discussion within each subreddit.
In this manner, Reddit provides a self-organising platform for online communities to develop and grow \citep{singer2014evolution}.

Activity on Reddit is \textit{self-exciting} in that each event on the platform increases the likelihood of subsequent events. 
This activity consists of users submitting titled \textit{posts} containing content that other users view, vote on, and discuss.
\textit{Comments} provide the basis for discussion where the sequence of comments on each post forms a network with a tree topology.
At the root of the \textit{discussion tree} is a node representing the initial post.
Each subsequent node represents a comment within the discussion, and nodes are linked by an edge when there is a \textit{reply-to} relation between them.
This reply-to relation links each comment to its \textit{parent} and defines the \textit{branching structure} of the discussion tree.
As such, discussions on Reddit unfold as a cascade of comments on a post.

In an analysis of all posts and comments submitted to Reddit from 2008 to 2014 inclusive (more than 150 million posts and 1.4 billion comments), \cite{medvedev2019modelling} demonstrated the utility of applying Hawkes' model for self-exciting point processes to online discussion \citep{hawkes1971point, hawkes1971spectra, hawkes1974cluster}.
Fitting these models to `small' (defined as those with between 50 and 200 nodes) and `larger' discussion trees (more than 200 nodes), \cite{medvedev2019modelling} obtained good predictions for the eventual size of a discussion, given an initial learning time interval of $s \in \left[ 4, 12 \right]$ hours.
\cite{krohn2019modelling} extended this approach to \textit{ex-ante} prediction for discussion trees, where a similarity graph of post titles and authors allowed predictions for new posts.

\begin{figure}[t]
    \centering
    \includegraphics[width=0.9\textwidth]{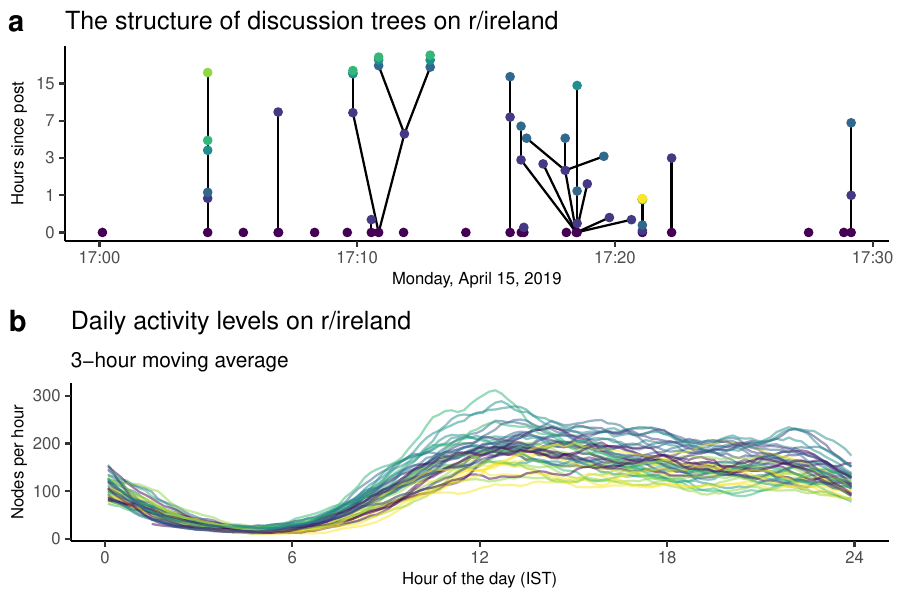}
    \caption{Plot \textbf{a} presents the discussion tree on the \texttt{r/ireland} subreddit over a $30$ minute period on Monday, April $15$, $2019$. Each node at 0 on the $y$-axis represents a post, while every other node corresponds to a comment. Lines represent reply-to relations between nodes. Many posts have few or no comments, while some prove popular and generate discussion. Plot \textbf{b} presents a 3-hour moving average for the number of posts and comments on the \texttt{r/ireland} subreddit. Each line represents a different day and each colour represents a specific day of the week. This figure illustrates the clear circadian rhythm in activity on the subreddit.
    }
    \label{fig:data_exploration}
\end{figure}

The Hawkes process offers a good model for online discussion dynamics; however, `off-the-shelf' implementations may fail to capture the idiosyncrasies of specific online communities.
For example, consider the discussions on the \texttt{r/ireland} subreddit from April 1 until May 12, 2019.
This data set consists of 117,787 nodes within 38,616 discussions.
Each node is associated with a unique identifier, a timestamp, and, where applicable, the identifier of its parent node. 
Exploratory data analysis identifies several salient features of this data set, apparently not considered by   \cite{medvedev2019modelling} or \cite{krohn2019modelling}.
Figure \ref{fig:data_exploration}\textbf{a} presents the structure of each discussion tree arising from posts submitted during 30 minutes on Monday, April 15, 2019. Of the 19 posts submitted, 11 have no replies, and the largest tree consists of 12 nodes.
Incidentally, none of these discussions meets the minimum size for inclusion in \cite{medvedev2019modelling}.
While \cite{krohn2019modelling} does address this issue, Figure \ref{fig:data_exploration}\textbf{b} highlights that activity within the subreddit follows a circadian rhythm, which is not accounted for within their model.
Activity on \texttt{r/ireland} tends to peak around 12:30 and is lowest between 04:00 and 06:00 Irish Summer Time (IST), suggesting that the discussion size may be time-dependent.
Models for online discussion should account for these features of empirical data.

\begin{figure}[t]
    \centering
    \includegraphics[width=0.9\textwidth]{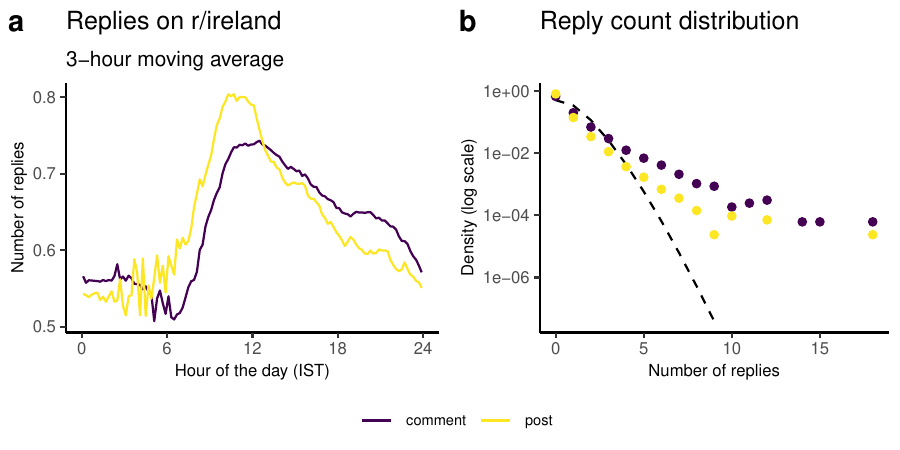}
    \caption{Plot \textbf{a} presents a 3-hour moving average number of replies to nodes on the \texttt{r/ireland} subreddit. This plot illustrates that the mean number of replies is lowest overnight and peaks around midday. Plot \textbf{b} illustrates the empirical distribution of the number of replies to a node. The dashed line represents the density of a Poisson distribution with a rate of 0.67, the overall mean number of offspring for nodes in our data set. This plot highlights the over-dispersed reply distributions observed in empirical data.}
    \label{fig:reply_exploration}
\end{figure}

Exploratory analyses allow us to identify several other features of the \texttt{r/ireland} data set. 
In particular, the link between each comment and its parent gives us the number of replies each node generates.
We explore this distribution in Figure \ref{fig:reply_exploration}.
Figure \ref{fig:reply_exploration}\textbf{a} highlights that the mean number of replies varies throughout the day, peaking at over 0.75 for nodes submitted around 11:00. 
In addition, Figure \ref{fig:reply_exploration}\textbf{b} demonstrates that the reply distribution has a heavy tail, suggesting that the distribution is over-dispersed.

\begin{figure}[t]
    \centering
    \includegraphics[width=0.9\textwidth]{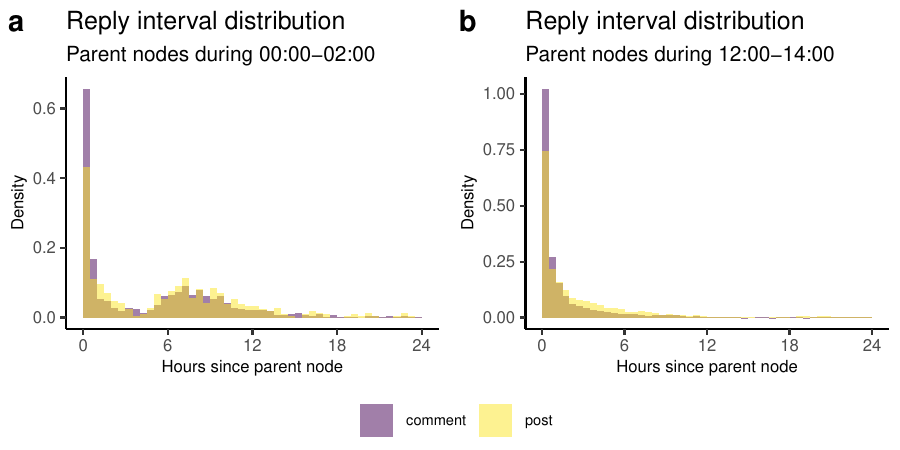}
    \caption{Plot \textbf{a} presents the reply interval distribution for nodes submitted between 00:00 and 02:00. 
We observe a bi-modal distribution such that nodes submitted in the very early hours of the morning are more likely to see a response between 08:00 and 10:00 than between 04:00 and 06:00. 
Plot \textbf{b} considers nodes submitted between 12:00 and 14:00. In this case, we observe a uni-modal distribution similar to that of an exponential random variable. Comparing these figures illustrates that the reply interval distribution within the \texttt{r/ireland} subreddit varies with the submission time of the parent node.}
    \label{fig:reply_interval_exploration}
\end{figure}

Finally, as reply-to relations also provide the time interval between each node and its children, we explore this distribution in Figure \ref{fig:reply_interval_exploration}. 
A comparison of reply intervals for nodes submitted at different times of day reveals stark differences in their respective distributions. 
For nodes submitted in the two hours after midday, the reply interval density decays at a roughly exponential rate; however, for nodes submitted just after midnight, we observe a bi-modal distribution. 
The second mode coincides with the hours of 08:00 and 10:00 in the morning when overall activity on the subreddit is relatively high, providing further evidence that the circadian rhythm of the \texttt{r/ireland} community influences discussion dynamics. 

Each feature identified in this exploratory analysis may represent an important consideration when modelling online discussion; however, standard Hawkes process typically models fail to account for such patterns.
Adapting the Hawkes process to include a circadian rhythm and over-dispersed reply distributions will allow for a more detailed analysis of discussions within this online community.

\subsection{Marked point processes}
\label{sec:point_processes}

Point processes provide a general framework for modelling discrete events occurring on some $d$-dimensional Euclidean space $\mathbb R^d$ \citep{daley2003introduction}. 
We are interested in a sequence of events over time, whereby event $i$ occurs at time $t_i \in \mathbb R$.
When events carry additional information, this information is referred to as a mark.
Marks occur on an arbitrary mark space $\mathcal K$, such that the mark $\kappa_i \in \mathcal K$ is associated with the event at $t_i$.
Thus, we define the \textit{marked point process} $\left\{ \left( t_i, \kappa_i \right) \right\}$ on $\mathbb R \times \mathcal K$ where the collection of times $\left\{ t_i \right\}$ is a point process known as the \textit{ground process}.

A marked point process is typically defined by its \textit{joint conditional intensity function}
\begin{equation}
    \lambda^* \left( t, \kappa \right) = \lambda^* \left( t \right) f^* \left( \kappa \mid t \right),
    \label{eq:joint_conditional_intensity}
\end{equation}
where $\lambda^* \left( t \right)$ is the conditional intensity of the ground process and $f^* \left( \kappa \mid t \right)$ is the conditional probability function of the mark $\kappa$ at time $t$. 
Note that the $^*$ notation serves as a reminder that these `functions' are themselves random variables which depend on $\mathcal F_{t^-}$, the filtration of the marked point process up to but not including $t$ (see Definition 7.3.III in \cite{daley2003introduction}). 

If we observe the marked point process $y = \left\{ \left( t_1, \kappa_1 \right), \dots \left( t_n, \kappa_n \right) \right\}$ on $\left[ 0, a \right) \times \mathcal K$, then by Proposition 7.3.III of \cite{daley2003introduction} the model likelihood is
\begin{equation}
        p \left( y \mid \theta \right) = \left[ \prod_{i = 1}^n \lambda^* \left( t_i \right) \right] \left[ \prod_{i = 1}^n f^* \left( \kappa_i \mid t_i \right) \right] \exp \left( - \Lambda^* \left( a \right) \right),
        \label{eq:joint_likelihood}
\end{equation}
where
\begin{equation*}
    \Lambda^* \left( a \right) = \int_0^a \lambda^* \left( u \right) du,
\end{equation*}
is the \textit{compensator} of the ground process, and $\theta$ is a set of parameters for the joint conditional intensity function.

Hawkes processes offer a particularly flexible class of point process models with an equivalent branching process representation \citep{hawkes1971point, hawkes1971spectra, hawkes1974cluster}.
The Hawkes process is typically defined as a self-exciting point process with a ground intensity function 
\begin{equation}
    \lambda^* \left( t \right) = \lambda_0 + \sum_{t_i < t} \gamma \left( t - t_i, \kappa_i \right),
    \label{eq:hawkes_intensity}
\end{equation}
where $\lambda_0 \in \mathbb R^+$ is referred to as the immigrant intensity and $\gamma \left( t - t_i, \kappa_i \right) \geq 0$ for all $t > t_i$ is the offspring intensity function.
This nomenclature is taken from the branching process view of the Hawkes process, where points are either immigrants or recursively generated offspring (see, e.g. \cite{rasmussen2013bayesian}).

\subsection{Epidemic dynamics and online activity}
\label{sec:epidemic_dynamics}

The dynamics underpinning the spread of content online are similar to those associated with infectious disease \citep{wallinga2004different, lloyd2005superspreading, bertozzi2020challenges, meagher2022assessing}. In this section, we draw analogies between these two areas so that modelling and inferential approaches in the epidemic literature can be usefully modified in the context of point process models for online discussions. 
Within the epidemic modelling framework, the point $t_i$ represents the time at which the $i$-th infection is recorded. Each infected individual is an \textit{index case}, giving rise to a random number of \textit{secondary infections}. We distinguish between two types of index case.
Those originating outside the population of interest are referred to as \textit{immigrants}, while those generated locally are \textit{offspring}.
Thus, models for the spread of infectious disease may apply to online discussions when each node in the discussion tree is analogous to an index case, posts are analogous to immigrants, and replies to posts are analogous to offspring. 
Moreover, by analogy, the number of replies corresponds to the number of secondary infections. 
As such, we refer to nodes giving rise to the most replies as the most \textit{infectious} nodes. 
These analogies draw a clear link between online discussion and epidemic dynamics.

Figure \ref{fig:reply_exploration}\textbf{b} illustrates that the reply distribution has a heavy tail, suggesting that some nodes generate many replies while the majority generate very few or none.
This is analogous to \textit{superspreading} in an epidemic, which emerges from heterogeneous disease reproduction \citep{lloyd2005superspreading}.
To model superspreading dynamics, consider the \textit{individual reproduction number} for index case (analogously, initial post) $i$, a latent variable denoted by $\nu_i > 0$, 
which governs the expected number of secondary infections (analogously, replies to the post) arising from that index case. 
Let $Z_i$ denote the number of secondary infections  arising from the index case at $t_i$ and assume that
\begin{equation}
    \begin{aligned}
        Z_i \mid \nu_i &\sim \pois \left( \nu_i \right), \\
        \nu_i &\sim \gam \left( \psi, \psi / \mu \right),
    \end{aligned}
    \label{eq:secondary_infections}
\end{equation}
where the shape $\psi$ and rate $\psi / \mu$ of the Gamma distributed $\nu_i$ are defined in terms of the \textit{reproduction number} $\mu$ and \textit{dispersion parameter} $\psi$. 
Integrating over $\nu_i$, we have that $Z_i \sim \nb \left( \mu, \psi \right)$ such that $\mathbb E \left[ Z_i \right] = \mu$ and $\operatorname{Var} \left( Z_i \right) = \mu + \mu^2 / \psi$.
As such, this hierarchical model permits an over-dispersed distribution for secondary infections.

One benefit of this modelling approach is that it allows us to estimate the expected proportion of secondary infections attributable to the most `infectious' index cases, as described by \cite{meagher2022assessing}.
If we let $f_\nu \left( x \right)$ and $F_\nu \left( x \right)$ denote the pdf and CDF for the Gamma distribution over individual reproduction numbers parameterised by $\mu$ and $\psi$, then the cumulative distribution function for the transmission of the disease is
\begin{equation}
    F_{\rep} \left( x \right) = \frac{1}{\mu} \int_{0}^{x} u \, f_\nu \left( u \right) du,
    \label{eq:transmission_quantiles}
\end{equation}
such that $F_{\rep} \left( x \right)$ is the expected proportion of index cases with $\nu_i < x$.
Thus, we can calculate, $r^{\psi}_{\alpha}$, the expected proportion of secondary infections attributable to the $100\alpha\%$ most infectious cases, where $\alpha\in(0,1)$, by finding $x_{\alpha}$ such that $1 - F_{\nu} \left( x_{\alpha}\right) = \alpha$, then $r^{\psi}_{\alpha} = 1 - F_{\rep} \left( x_{\alpha} \right)$. In practice, $r^{\psi}_{\alpha}$ is estimated numerically, as it is not available in closed form. 
This analysis provides a straightforward interpretation of the dispersion parameter $\psi$. We refer the reader to Section $3.1.2$ of \cite{meagher2022assessing} for further details of the calculation described here. 

Adapting the model for superspreading epidemic dynamics defined by (\ref{eq:secondary_infections}) to online discussion, as set out in Section~\ref{sec:generative_model}, will provide new insight into discussion dynamics. In addition, the identity in (\ref{eq:transmission_quantiles}) both guides the choice of prior for $\psi$ in Section~\ref{sec:parameter_inference} and provides a useful interpretation of its posterior uncertainty, as outlined in Section~\ref{sec:inference}.

\section{Methods}
\label{sec:methods}

\subsection{The generative model}
\label{sec:generative_model}

In order to define our generative model for discussion trees, we start with a single discussion. 
Following Section~\ref{sec:epidemic_dynamics}, we refer to the discussion tree as a \textit{cluster}, nodes as \textit{points}, posts as \textit{immigrants}, replies as \textit{offspring}, and the collection of reply-to relations linking each point to its parent as the \textit{branching structure}.
We also refer to a \textit{reproduction number} for each point, which is proportional to its expected number of offspring, while the time between a point and its parent is the \textit{generation interval}.

Consider the cluster $\boldsymbol y_{t_n} = \left( \boldsymbol  t, \boldsymbol  \beta \right)$ where $\boldsymbol y_{t_n}$ consists of $n$ points at times $\boldsymbol t = \left(t_{1}, \dots, t_{n} \right)^\prime$ and a branching structure $\boldsymbol \beta = \left(\beta_{1}, \dots, \beta_{n} \right)^\prime$.
This cluster consists of two types of points, whereby the immigrant at $t_1$ seeds the cluster and all remaining points are offspring.
The branching structure links each point to its parent via an edge $\beta_k = j$ if $t_j$ is the parent of $t_k$. 
The branching structure is complete given $\beta_1 = 0$.
Thus, when $t_{j-1} < t_j$ we have $t_{j} \in \left(t_{1}, a \right]$, and $\beta_{j} \in \left\{1, \dots, j-1 \right\}$ for all $j = 2, \dots, n$. 

Our objective is to model the trajectory of the ground process $\boldsymbol t$.
As such, we treat the immigrant at $t_1$ as a fixed point seeding the cluster and include only the offspring $t_2, \dots, t_n$ in the generative model.
In addition, we introduce the set of latent marks $\boldsymbol \nu = \left(\nu_{1}, \dots, \nu_{n} \right)^\prime$ such that $\nu_{j} > 0$ for all $j = 1, \dots, n$.
Here, $\nu_j$ is the individual reproduction number representing unknown factors influencing the expected number of offspring from point $j$.
Our generative model is defined as follows.

\begin{model}[A generative model for online discussion trees] \label{mod:generative_model}
\hspace{1pt}
\begin{enumerate}
    \item The immigrant $t_{1}$ is generated by some exogenous process. We set $\beta_{1} = 0$ and generate the individual reproduction number $\nu_{1}$, which follows the conditional mark density $f^* \left( \nu_{1} \mid t_1, \beta_1 \right)$.
    \item The immigrant $t_{1}$ generates the cluster $\boldsymbol y_{t_n}$, where $\boldsymbol y_{t_n}$ is a collection of points in generations $l = 0, 1, 2, \dots$ with the following branching structure: 
    \begin{enumerate}
        \item Generation 0 consists of the immigrant $\left( t_{1}, \beta_{1}, \nu_{1} \right)$ while offspring in the subsequent generations are generated recursively. 
        \item Given the $0, \dots, l$ generations of $\boldsymbol y_{t_n}$, each $\left( t_{j}, \beta_{j}, \nu_{j} \right)$ in generation $l$ generates a Poisson process $\Phi_{j}$ of offspring in generation $l+1$ with intensity $\gamma_{j} \left( t \right) = \gamma \left( t \mid t_{j}, \beta_{j}, \nu_{j} \right) \geq 0$ defined for $t > t_{j}$. We refer to $\Phi_{j}$ as the offspring process and $\gamma_{j} \left( \cdot \right)$ as the offspring intensity. \label{mod:recursion}
        \item Each offspring $t_{k} \in \Phi_{j}$ has an associated edge $\beta_{k} = j$ and individual reproduction number $\nu_{k}$ with conditional mark density $f^* \left( \nu_{k} \mid t_k, \beta_k \right)$.
    \end{enumerate}
\end{enumerate}
\end{model}

This generative model provides a flexible framework for modelling self-exciting processes. 
Its ground intensity function is of the form
\begin{equation}
    \lambda^* \left( t \right) = \sum_{t_j < t} \gamma_j \left( t \right),
    \label{eq:generative_model_intensity}
\end{equation}
which is similar to that of the Hawkes process presented in (\ref{eq:hawkes_intensity}); however, in our formulation, we explicitly condition $\gamma_j \left( t \right)$ on the time, position within the branching structure and the latent reproduction number associated with point $j$.
Similarly, the conditional mark density $f^* \left( \nu_{j} \mid t_j, \beta_j \right)$ depends on the branching structure.
Such conditioning offers enormous flexibility to specify models that capture essential features of our data while including the standard Hawkes process as a special case.

Given that the exploratory analysis in Section \ref{sec:reddit} identified a circadian rhythm and over-dispersed offspring distributions within discussions on the \texttt{r/ireland} subreddit, we propose the following functional form for our generative model. 
We specify the offspring intensity function from the point at $t_j$ as
\begin{equation}
    \gamma_j \left( t \right) = \nu_j  \, \alpha \left( t \right) \rho \left( t - t_j \mid \beta_j \right),
    \label{eq:offspring_intensity}
\end{equation}
where the reproduction number $\nu_j$ scales the offspring intensity such that points with large reproduction numbers are more likely to produce offspring, the \textit{activity function} $\alpha \left( t \right) > 0$ models the circadian rhythm of activity on the board, and the \textit{excitation function} $\rho \left( u \mid \beta_j \right) > 0$ defined for $u > 0$ such that $\int_0^{\infty} \rho \left( u \mid \beta_j \right) du = 1$ models the distribution of generation intervals between offspring and their parent $t_j$. 

We assume that the activity function $\alpha \left( t \right)$ is a periodic function with a known period $T$.
Since $\alpha \left( t \right)$ models a circadian rhythm, the period $T$ corresponds to 24 hours.
In addition, we assume that $\alpha \left( t \right)$ is normalised such that $\int_{0}^T \alpha \left( u \right) du = T$.
This normalisation, which implies that the mean of activity function over the period is one, ensures that each element within the offspring intensity function is identifiable.
To satisfy these constraints, we define $\alpha \left( t \right)$ via sinusoidal basis functions such that
\begin{equation}
    \alpha \left( t \right)
    =\alpha_0 + \sum_{k = 1}^K \left( \alpha_{2k-1} \sin \left( \omega_k t \right) + \alpha_{2k} \cos \left( \omega_k t \right) \right)
    = \boldsymbol \alpha^\prime \boldsymbol s_{\boldsymbol \omega} \left( t \right)
    \label{eq:activity_function}
\end{equation}
where we have $2K + 1$ real-valued sinusoidal coefficients $\boldsymbol \alpha = \left(\alpha_0, \alpha_1, \dots, \alpha_{2K} \right)^\prime$ and sinusoidal basis functions $\mathbf{s}_{\boldsymbol \omega} \left( t \right) = \left( s_0, s_1 \left( t \right), \dots, s_{2K} \left( t \right) \right)^\prime$ for $K$ non-negative basis frequencies $\boldsymbol \omega = \left( \omega_1, \dots, \omega_K \right)^\prime$ such that $\alpha_0 = s_0 = 1$ and where we define
\begin{equation*}
    \begin{aligned}
    s_{2k - 1} \left( t \right) &= \sin \left( \omega_k t \right), \hspace{1cm}  
    &s_{2k} \left( t \right) &= \cos \left( \omega_k t \right)
    \end{aligned}
\end{equation*}
for $k = 1, \dots, K$. 
Given that $\omega_k = 2 \pi f_k$ where $f_k$ is the number of oscillations of the sinusoidal function per period $T$, selecting $f_k = x / T$ such that $x \in \mathbb N$ for all $k$ normalises $\alpha \left( \cdot \right)$, as required.
Note that the generative model is misspecified if $\alpha \left( t \right) < 0$ for any $t$, i.e., we cannot have a negative offspring intensity, and so care must be taken when specifying $\boldsymbol \alpha$. 
This issue can be addressed in a Bayesian inference scheme by specifying a shrinkage prior for $\boldsymbol \alpha$.

Note that we condition the excitation function $\rho \left( u \mid \beta_j \right)$ on the position of $t_j$ within the branching structure, which allows the excitation associated with immigrants and offspring to differ.
In the context of online discussions, this is a crucial distinction as immigrants and offspring tend to present different empirical offspring distributions \citep{medvedev2019modelling, gleeson2020branching}. 
We might expect these differences as immigrants represent posts to the online board, while the offspring are comments that only occur after a user has viewed the content within the post.
To impose this conditioning while maintaining a relatively simple model, we adopt an exponential excitation function
\begin{equation}
    \rho \left( t - t_j \mid \beta_j \right) = 
    \begin{cases}
    \e \left(t - t_j \mid \eta_1 \right), & \text{if } \beta_j = 0, \\
    \e \left(t - t_j \mid \eta_2 \right), & \text{otherwise},
    \end{cases}
    \label{eq:excitation_function}
\end{equation}
such that the excitation function follows the pdf of an exponential random variable given decay rates $\boldsymbol \eta = \left( \eta_1, \eta_2 \right)^\prime$.
That is to say, the rate of excitation decay for immigrants is $\eta_1$, while for offspring, it is $\eta_2$.

All that remains is to specify the conditional mark density for the individual reproduction number $\nu_j$. 
We adopt the approach to modelling epidemic superspreading proposed by \cite{lloyd2005superspreading} and assume that individual reproduction numbers follow a Gamma distribution. 
In addition, we allow the distributions for individual reproduction numbers for immigrants and offspring to differ such that 
\begin{equation}
    f^* \left( \nu_j \mid t_j, \beta_j \right) =
    \begin{cases}
    \gam \left(\nu_j \mid \psi_1, \psi_1 / \mu_1 \right), & \text{if } \beta_j = 0, \\
    \gam \left(\nu_j \mid \psi_2, \psi_2 / \mu_2 \right), & \text{otherwise}.
    \end{cases}
    \label{eq:reproduction_number_density}
\end{equation}
Here, we define the shape and rate of the Gamma distributions in terms of the immigrant and offspring reproduction numbers $\boldsymbol \mu = \left( \mu_1, \mu_2 \right)^\prime$ and dispersion parameters $\boldsymbol \psi = \left( \psi_1, \psi_2 \right)^\prime$, respectively.
This implies that $\mathbb E \left[ \nu_1 \right] = \mu_1$ and $\var \left( \nu_1 \right) = \mu_1^2 / \psi_1$ while $\mathbb E \left[ \nu_j \right] = \mu_2$ and $\var \left( \nu_j \right) = \mu_2^2 / \psi_2$ for $j > 1$, allowing us to model over-dispersed offspring distributions.

\begin{figure}[ht]
    \centering
    \includegraphics[width=0.9\textwidth]{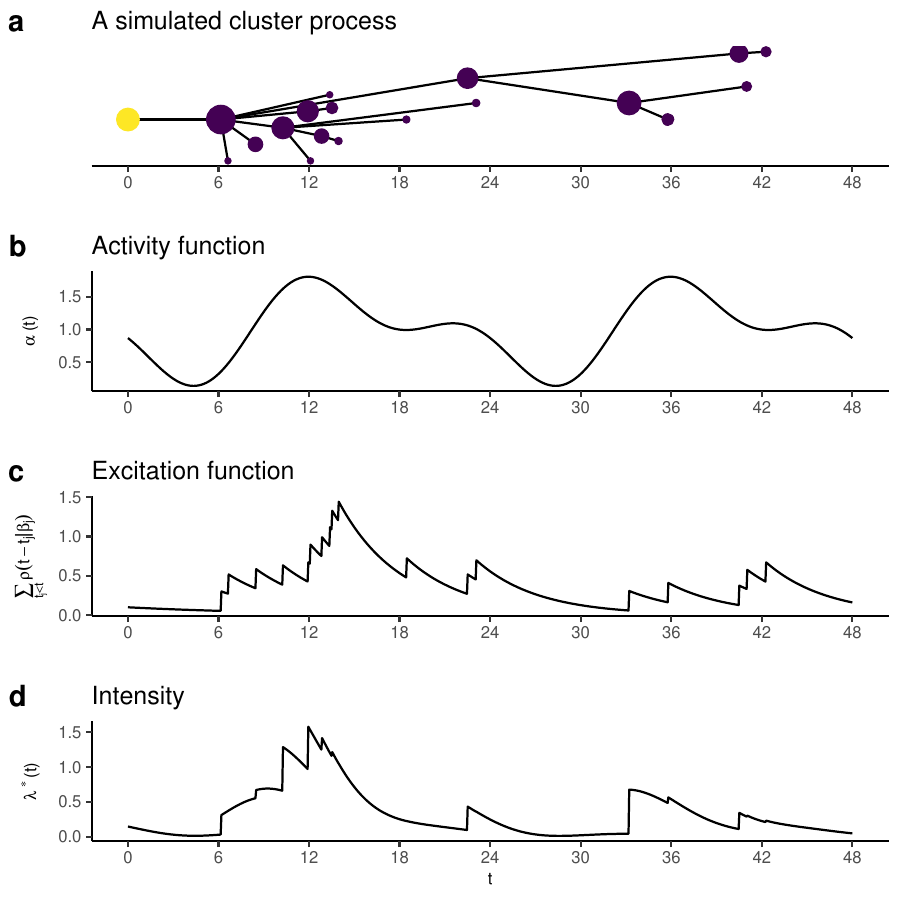}
    \caption{An illustration of the generative model presented in Section \ref{sec:generative_model}.
    Plot \textbf{a} presents a simulated cluster such that the immigrant point is yellow, offspring points are purple, and the size of each point is proportional to its individual reproduction number. Line segments link each offspring node to its parent. Plot \textbf{b} is the activity function $\alpha \left( t \right)$, modelling a circadian rhythm. Note that $\alpha \left( t \right)$ is a periodic function over 24 hours. In plot \textbf{c}, we see the cumulative sum of the excitation functions induced by points up to time $t$. Note that this cumulative sum does not include the effect of individual reproduction numbers. Finally, \textbf{b} illustrates the conditional ground intensity function for the marked point process obtained by substituting equations (\ref{eq:excitation_function}) and (\ref{eq:activity_function}) into (\ref{eq:offspring_intensity}), which then substitutes into (\ref{eq:generative_model_intensity}).}
    \label{fig:generative_model}
\end{figure}

We have defined a generative model for online discussion trees, as illustrated in Figure \ref{fig:generative_model}.
Note that it is straightforward to specify simplified versions of this model.
For example, if we do not wish to include a circadian rhythm, then we specify $\boldsymbol \omega = \emptyset$ such that $\alpha (t) = 1$ for all $t$.
Similarly, if the heterogeneity introduced by individual reproduction numbers $\nu_j$ is inappropriate, then we set $\nu_1 = \mu_1$ and $\nu_j = \mu_2$ for all $j = 2, \dots, n$.
This flexibility is a characteristic of Hawkes-type branching processes, allowing us to specify models that capture features of empirical data.

\subsubsection{Likelihood}
\label{sec:likelihood}

The generative model in Section \ref{sec:generative_model} describes a marked point process. As such, it has a joint conditional intensity function and likelihood of the form presented in equations (\ref{eq:joint_conditional_intensity}) and (\ref{eq:joint_likelihood}), respectively.
Thus, we define the likelihood of our model as follows.

In general, the ground intensity for a Hawkes process is of the form presented in (\ref{eq:hawkes_intensity}), and computing the intensity associated with the point at $t_j$ scales with $\mathcal O \left( n^2 \right)$. 
In this context, however, the underlying branching structure is known, reducing the intensity associated with $t_j$ to the offspring intensity function for its parent $t_{\beta_j}$. As such, we have that
\begin{equation}
    \lambda^* \left( t_j \right) = \gamma_{\beta_j} \left( t_j \right) =  \nu_{\beta_j} \alpha \left( t_j \right) \rho \left( t_j - t_{\beta_j} \mid \beta_{\beta_j} \right).
    \label{eq:conditional_intensity}
\end{equation}
Substituting (\ref{eq:activity_function}) and (\ref{eq:excitation_function}) into (\ref{eq:conditional_intensity}) and then substituting (\ref{eq:conditional_intensity}) and (\ref{eq:reproduction_number_density}) into (\ref{eq:joint_conditional_intensity}) provides the joint conditional intensity function for our generative model. 

In order to derive an expression for the model likelihood, we first treat the sinusoidal basis frequencies $\boldsymbol \omega$ as fixed hyper-parameters of the model. 
Thus, we are left with parameters $\theta = \left\{ \boldsymbol \alpha, \boldsymbol \eta, \boldsymbol \mu, \boldsymbol \psi \right\}$ such that the complete-data likelihood for the cluster $\boldsymbol y_{t_n}$ and latent variables $\boldsymbol \nu$ is of the form
\begin{equation}
\begin{aligned}
    p \left(\boldsymbol y_{t_n}, \boldsymbol \nu \mid \theta \right) 
    &= p \left(\boldsymbol t \mid \boldsymbol \beta, \boldsymbol \nu, \boldsymbol \alpha, \boldsymbol \eta \right) p \left(\boldsymbol \nu \mid \boldsymbol \beta, \boldsymbol \mu, \boldsymbol \psi \right) \\
    &= \left[ \prod_{j = 2}^n \lambda^* \left( t_j \right) \right] \exp \left( - \Lambda^* \left( a \right)\right)  \left[ \prod_{j = 1}^n f^* \left( \nu_j \mid \beta_j \right) \right],
\end{aligned}
\label{eq:complete_likelihood}
\end{equation}
where we have dropped the conditioning on $\boldsymbol \omega$ for a more parsimonious notation.
We have an analytic expression for the compensator (see Appendix \ref{app:compensator_derivation}) such that
\begin{equation}
    \Lambda^* \left( a \right) = \nu_{1} \, \boldsymbol \alpha^\prime \boldsymbol W \left(t_{1}, a, \eta_1 \right) + \sum_{j = 2}^{n} \nu_{j} \, \boldsymbol \alpha^\prime \boldsymbol W \left(t_{j}, a, \eta_2 \right),
\label{eq:compensator}
\end{equation}
where we have the $2K + 1$ dimensional vector $\boldsymbol W \left(t, a, \eta \right)$ with elements
\begin{equation*}
    W_k \left(t, a, \eta \right) = \int_{t}^a s_k \left( u \right) \, \eta \exp \left( - \eta \left( u - t \right) \right) du,
\end{equation*}
for $k = 0, 1, \dots, 2K$.
Finally, integrating over the latent marks $\boldsymbol \nu$ (see Appendix \ref{app:integration}) the likelihood for our generative model is
\begin{equation}
    \begin{aligned}
        p \left( \boldsymbol y_{t_n} \mid \theta \right)
        &=  \int  p \left(\boldsymbol y_{t_n}, \boldsymbol \nu \mid \theta \right) d \boldsymbol \nu, \\
        &=  \frac{\Gamma \left( \psi_1 + z_{1} \right)}{\Gamma \left( \psi_1 \right)} 
        \left( \frac{\mu_1}{\psi_1 + \mu_1 c_{t_{1}}} \right)^{z_{1}}
        \left( \frac{\psi_1}{ \psi_1 + \mu_1 c_{t_{1}}} \right)^{\psi_1} \\
        & \hspace{0.5cm}  \left[ \prod_{j = 2}^{n} \alpha \left( t_{j} \right) \rho \left( t_{j} - t_{\beta_{j}} \mid \beta_{j} \right) 
        \frac{\Gamma \left( \psi_2 + z_{j} \right)}{\Gamma \left( \psi_2 \right)} 
        \left( \frac{\mu_2}{\psi_2 + \mu_2 c_{t_{j}}} \right)^{z_{j}}
        \left( \frac{\psi_2}{ \psi_2 + \mu_2 c_{t_{j}}} \right)^{\psi_2}
        \right], 
    \end{aligned}
    \label{eq:model_likelihood}
\end{equation}
where $z_{j} = \sum_{k = 1}^n \delta \left( \beta_k = j \right)$ is the observed number offspring from point $j$ and we have introduced $c_{t_{1}} = \boldsymbol \alpha^\prime \boldsymbol W \left( t_{1}, a, \eta_1 \right)$ and $c_{t_{j}} = \boldsymbol \alpha^\prime \boldsymbol W \left( t_{j}, a, \eta_2 \right)$, for $j = 2, \dots, n$ to simplify notation.
Note that $\delta \left( \cdot \right)$ is the usual indicator function such that $\delta \left( x \right) = 1$ when the condition $x$ is true and $\delta \left( x \right) = 0$ otherwise.
Thus, we have a model likelihood that scales with $\mathcal O \left( n \right)$ rather than the $\mathcal O \left( n^2 \right)$ typically associated with self-exciting processes, making it possible to develop efficient likelihood-based inference schemes for this generative model. 

\subsubsection{Simulation}
\label{sec:simulating}

It is straightforward to simulate from our generative model using an algorithm similar to that described by \cite{moller2006approximate}, where we recursively simulate the Poisson processes described in point \ref{mod:recursion} of Model \ref{mod:generative_model}. 
Thinning algorithms provide a general approach to simulating Poisson processes (e.g., Algorithm 7.5.II or Algorithm 7.5.IV in \cite{daley2003introduction}). We outline our simulation algorithm as follows.

Consider the cluster of points $\boldsymbol y_{t_n}$ on the interval $\left[ t_1, a \right)$ for $a \geq t_1$. 
Note that we permit the case where $a = t_1$, which implies that only the immigrant point is known.
Our objective is to simulate a realisation $\boldsymbol y_{t_{n + m}}^* = \left( \boldsymbol t^*, \boldsymbol \beta^* \right)$ which propagates the cluster $\boldsymbol y_{t_n}$ over the interval $\left[ a, b \right)$ given parameters $\theta$ and hyper-parameters $\boldsymbol \omega$. 
Here, we have $\boldsymbol t^* = \left( t_1^*, \dots, t_{n+m}^*\right)^\prime$ and $\boldsymbol \beta^* = \left( \beta_1^*, \dots, \beta_{n+m}^*\right)^\prime$ where $\boldsymbol t_{1:n}^* = \boldsymbol t$, $\boldsymbol \beta_{1:n}^* = \boldsymbol \beta$ and $t_{n + j}^* \in \left[ a, b \right)$, $\beta_{n+j}^* \in \left\{1, \dots, n+j-1 \right\}$ for $j = 1, \dots, m$.
That is to say, the first $n$ points in $\boldsymbol y_{t_{n + m}}^*$ are equivalent to those in $\boldsymbol y_{t_{n}}$, while the remaining $m = 0, 1, 2, \dots$ points represent a simulated realisation from the generative model. 
In addition, we introduce $\boldsymbol \nu^* = \left( \nu_1^*, \dots, \nu_{n+m}^* \right)$, the latent individual reproduction numbers associated with each of the points in $\boldsymbol y_{n+m}^*$.

Our first step is to obtain a posterior distribution for $\boldsymbol \nu_{1:n}^*$, the reproduction numbers for the points at $\boldsymbol t$.
We show in Appendix \ref{app:integration} that the posterior distribution for each $\nu_j$ within our generative model is of the form
\begin{equation}
     p \left( \nu_j \mid \boldsymbol y_{t_n}, \theta \right) 
     = \gam \left(\nu_j \mid z_j + \psi, \boldsymbol \alpha^\prime \boldsymbol W \left(t_1, a, \eta \right) + \frac{\psi}{\mu} \right),
\label{eq:fitness_posterior}
\end{equation}
where $\eta, \mu$, and $\psi$ are the corresponding memory decay rate, reproduction number, and dispersion parameter, respectively.
With this expression in place, we sample $\boldsymbol \nu_{1:n}^*$ and recursively generate points on the interval $\left[a, b \right)$ according to \ref{mod:recursion} in Model \ref{mod:generative_model}, given an algorithm simulating the Poisson process $\Phi_j$ for $j = 1, 2, \dots$.

We propose a thinning algorithm for simulating $\Phi_j$. 
In this approach, we first generate a realisation of the Poisson process $\tilde \Phi_j$ over $\left[ a, b \right)$ with intensity $\tilde \gamma_j \left( u \right) \geq \gamma_j \left( u \right)$ for all $u \in \mathbb R^+$.
We then assign each $t_k \in \tilde \Phi_j$ to $\Phi_j$ with probability $\gamma_j \left( t_k \right) / \tilde \gamma_j \left( t_k \right)$.
Thus, we generate the realisation $\boldsymbol y_{t_{n + m}}^*$, as required.
We include a more detailed description of the simulation algorithm in Appendix \ref{app:simulation_algorithm}.

\subsection{Prior modelling and parameter inference}
\label{sec:parameter_inference}

Consider a training data set $\boldsymbol y_{\tr} = \left\{ \boldsymbol y_{t_{n}}^{(1)}, \dots, \boldsymbol y_{t_{n}}^{(S)} \right\}$ where the superscript $i$ notation indicates that we have the cluster $\boldsymbol y_{t_{n}}^{i} = \left(\boldsymbol t_i, \boldsymbol \beta_i \right)$ of $n_i$ nodes at times $t_{ij} \in \left[t_{i1}, a_i \right)$ with edges $\beta_{ij} \in \left\{0, 1, \dots, j-1 \right\}$ for $j = 1, \dots, n_i$ and $i = 1, \dots, S$. 
We assume that each of the $S$ clusters is an independent realisation from our generative model parameterised by $\theta$ and $\boldsymbol \omega$. 
Thus, the likelihood for each cluster is of the form outlined in Section \ref{sec:likelihood} and the overall likelihood is the product of the $S$ cluster likelihoods.

In our Bayesian approach to parameter inference, we draw Markov Chain Monte Carlo (MCMC) samples from the posterior density
\begin{equation}
    p \left( \theta \mid \boldsymbol y_{\tr}  \right) 
    \propto \prod_{i = 1}^S  \bigg[ p \left( \boldsymbol y_{t_{n}}^{i}  \mid \theta \right) \bigg] p \left( \theta \right),
    \label{eq:posterior_definition}
\end{equation}
where $p \left( \boldsymbol y_{t_{n}}^{i}  \mid \theta\right)$ is the likelihood defined by (\ref{eq:model_likelihood}) and $p \left( \theta \right)$ is our prior distribution for the model parameters. We define this prior as follows.

Firstly, we consider the sinusoidal coefficients $\boldsymbol \alpha$, which define the activity function modelling the circadian rhythm of $\gamma_{ij} \left( t \right)$. 
Given that $\alpha_0 = 1$ and $\alpha_k \in \mathbb R$ for all $k = 1, \dots, 2K$, we assume that 
\begin{equation}
    \alpha_k \sim \norm \left( 0, \sigma_\alpha^2 \right).
    \label{eq:alpha_prior}
\end{equation}
To specify $\sigma_\alpha$, consider the following.
The sinusoidal basis functions, which we have defined in terms of their Cartesian coordinates, have an equivalent polar coordinate form such that
\begin{equation}
    \alpha \left(t\right) 
    = 1 + \sum_{k = 1}^K A_{k} \cos \left( \omega_k t + \varphi_k \right),
\end{equation}
with amplitude $A_k = \sqrt{\alpha_{1, k}^2 + \alpha_{2, k}^2}$ and phase $\varphi_k = \arctan\left(\alpha_{1, k}, \alpha_{2, k}\right)$. The polar form is more straightforwardly related to the $\alpha \left( t \right) > 0$ constraint. While this constraint may be satisfied when $\sum_{k = 1}^K A_k > 1$ if $K > 1$ and the phase of distinct sinusoids dampens the negative oscillation of $\alpha \left( t \right)$, we cannot have $\sum_{k = 1}^K A_k \gg 1$. As such, it is fair to assume that $\sum_{k = 1}^K\mathbb E \left[ A_k \right] \leq 1$, which, given our iid normal prior for $\alpha_k$, implies that $2K \mathbb E \left[ \alpha_k^2 \right] \leq 1$. Thus, we specify $\sigma_\alpha = 1 / \sqrt{2K}$, which maximises the entropy of our prior within this constraint.
This prior acts as a shrinkage prior on $\boldsymbol \alpha$, whereby the degree of shrinkage towards 0 is proportional to $K$. In practice, this allows us to avoid specifying a model with a negative conditional intensity function.

Next, we consider the decay rate for the immigrant offspring process $\eta_1$. 
In the absence of any exogenous effects (i.e. when $\alpha \left( t \right) = 1$), this parameter defines the distribution of generation intervals between the immigrant and its children, such that the expected generation interval is $1 / \eta_1$. 
There is little prior information on the expected generation interval when modelling online discussion. 
Discussions may occur over minutes, days, weeks or even months.
As such, specifying an expected generation interval, \textit{a priori} is challenging. 
A Gamma prior for $\eta_1$ with a shape parameter of 1 or less is an appealing option.
This prior implies that the expected inter-arrival time follows an Inverse-Gamma distribution for which the expectation is undefined.
The same argument holds for the offspring decay rate $\eta_2$, and so, for $l = 1, 2$, we set
\begin{equation}
    \eta_l \sim \gam \left(1, 1 \right).
    \label{eq:eta_prior}
\end{equation}

The permissible range of reproduction numbers $\boldsymbol \mu$ is well defined, \textit{a priori}.
If we have $\mu_1 > 0$ and $\mu_2 > 1$, then $\mathbb E \left[ n_i \right] \to \infty$ as $a \to \infty$.
Because we only ever observe $\boldsymbol y_{t_{n}}^{i} $ on a finite interval, it is possible to have $\mu_2 > 1$ and, because there is only ever one immigrant per cluster, it is also possible that $\mu_1 > 1$. 
However, we view both possibilities as highly unlikely given that the explosive growth they imply would be subject to physical constraints set by the infrastructure supporting the online board. 
We expect that $\mu_1, \mu_2 \in \left( 0, 1 \right)$.
Furthermore, these reproduction numbers are unlikely to be very close to zero or one, as this would indicate that clusters should all be very small or very large, respectively.
Thus, we specify a Gamma prior with an expectation of $0.5$ and standard deviation of $0.25$, such that our prior on $\mu_l$ for $l = 1, 2$ is
\begin{equation}
    \mu_l \sim \gam \left(4, 8 \right).
    \label{eq:mu_prior}
\end{equation}

Finally, we consider the dispersion parameters $\psi$. Following Section~\ref{sec:epidemic_dynamics} and as described by \cite{meagher2022assessing}, we link the value of these dispersion parameters to the expected proportion of offspring arising from the most `infectious' points.
For example, if we expect that the 20\% most infectious points give rise to 30-90\% of total offspring, so that $r_{0.2}^{\psi_i} \in (0.3,0.9),\; i=1,2$, then it is likely that $\psi_1, \psi_2 \in \left(0.1, 10 \right)$.
This assumption, which seems reasonable \textit{a priori}, can be encoded as a log-normal prior. As such, we specify
\begin{equation}
    \ln \psi_l \sim \norm \left( 0, 1 \right),
    \label{eq:psi_prior}
\end{equation}
for $l = 1, 2$. 
Note that $\psi_l > 10$ would suggest that individual reproduction numbers are relatively homogeneous. 
If we wish to model such an offspring process, then a more parsimonious approach is to assume that $\psi_l \to \infty$. In this case, the number of offspring from each point is Poisson distributed, and the model likelihood is of the form
\begin{equation}
        p \left( \boldsymbol y_{t_{n}}^{i} \mid \theta \right)
        =  \mu_1^{z_{i1}} \exp \left( - \mu_1 c_{t_{i1}} \right) 
        \left[ \prod_{j = 2}^{n_i} \alpha \left( t_{ij} \right) \rho \left( t_{ij} - t_{i\beta_{ij}} \mid \beta_{ij} \right) 
        \mu_2^{z_{ij}} \exp \left( - \mu_2 c_{t_{ij}} \right) 
        \right].
    \label{eq:homogeneous_likelihood}
\end{equation}

With that, we have fully specified $p \left( \theta \right)$ and turn our attention to the MCMC sampling scheme.
To sample from (\ref{eq:posterior_definition}), we code our model as a Stan program \citep{carpenter2017stan}.
Given the likelihood (\ref{eq:model_likelihood}), we define the joint posterior density over model parameters as
\begin{equation}
    p \left( \theta \mid \boldsymbol y_{\tr}  \right)
    \propto \left[ \prod_{i = 1}^S p \left( \boldsymbol y_{t_{n}}^{i} \mid \theta \right) \right] \left[ \prod_{k = 1}^{2K} p \left( \alpha_k \mid K \right) \right] \left[ \prod_{l = 1}^{2} p \left( \eta_l \right) p \left( \mu_l \right) p \left( \psi_l \right) \right],
\label{eq:stan}
\end{equation}
where and $ p \left( \alpha_k \mid K \right), p \left( \eta_l \right), p \left( \mu_l \right)$, $p \left( \psi_l \right)$ follow from (\ref{eq:alpha_prior}), (\ref{eq:eta_prior}), (\ref{eq:mu_prior}), and (\ref{eq:psi_prior}) above. In the special case where we let $\psi_l \to \infty$, we substitute in (\ref{eq:homogeneous_likelihood}) for the likelihood and remove the relevant dispersion parameter prior. Code implementing this sampling scheme is available in an R package at \href{https://github.com/jpmeagher/onlineMessageboardActivity}{https://github.com/jpmeagher/onlineMessageboardActivity}.

\subsection{Model Assessment}
\label{sec:model_assessment}

\subsubsection{Model evidence}

Consider a particular specification of our generative model, denoted $\mathcal M$, with parameters $\theta$. 
The evidence for model $\mathcal M$, denoted $\mathcal E$, is the normalising constant of the posterior distribution, that is, the density of the data under model $\mathcal M$. 
In this context, we have
\begin{equation}
    \mathcal E = p \left( \boldsymbol y_{\tr} \mid \mathcal M  \right) = \int \prod_{i = 1}^S p \left( \boldsymbol y_{t_{n}}^{i}   \mid \theta, \mathcal M  \right) p \left( \theta \mid \mathcal M \right) d \theta,
\end{equation}
where we explicitly condition the likelihood and prior on $\mathcal M$.
This integral is intractable and so we must estimate $\mathcal E$.
Many approaches for estimating the model evidence are available (see, e.g. \cite{friel2012estimating} for a review). 
Here, we take the bridge-sampling approach proposed by \cite{meng1996simulating} and implemented in the \texttt{bridgesampling} R package \citep{gronau2020package}. 
Details are presented by \cite{gronau2017tutorial}.

To compare the candidate models $\mathcal M_l$ and $\mathcal M_{l^\prime}$ for $l \neq l^\prime$, we estimate the Bayes factor
\begin{equation}
    \mathcal{BF}_{ll^\prime} = \mathcal E_l / \mathcal E_{l^\prime},
\end{equation}
where $\mathcal E_l$ and $\mathcal E_{l^\prime}$ is the evidence for $\mathcal M_l$ and $\mathcal M_{l^\prime}$, respectively.
This ratio allows us to assess the evidence supporting one model over another.
For, example, finding $\ln \mathcal{BF}_{l l^\prime} > 2.3$ represents strong evidence in favour of $\mathcal M_l$ over $\mathcal M_{l^\prime}$ \citep{kass1995bayes}.

\subsubsection{Predictive performance}

We assess the predictive performance of $\mathcal M_l$ given a test set $\boldsymbol y_{\ts}$ of $S^\prime$ clusters where $\boldsymbol y_{t_{n}}^{i} \in \boldsymbol y_{\ts}$ consists of points $t_{ij} \in \left[ t_{i1}, a_i\right)$ and we have samples $\theta_l^{\left( 1 \right)}, \dots, \theta_l^{\left( R \right)}$ from $p \left( \theta \mid \boldsymbol y_{\tr}, \mathcal M_l \right)$.
We consider two approaches to assessing predictive performance based on a proper scoring rules \citep{gneiting2007strictly}.

In the first case, we estimated the expected log cluster-wise predictive density of $\mathcal M_l$ on the test set such that
\begin{equation}
    \widehat{\lpd}_l = \sum_{i = 1}^{S^\prime} \ln \left( \frac{1}{R} \sum_{r = 1}^R p \left( \boldsymbol y_{t_{n}}^{i} \mid \theta_l^{\left( r \right)}, \mathcal M_l \right) \right).
\end{equation}
In this setting, larger values for $\widehat{\lpd}_l$ indicate better predictive performance.
We compare candidate models via
\begin{equation}
    \Delta \widehat{\lpd}_{l l^\prime} = \sum_{i = 1}^{S^\prime} \left[ \ln \left( \frac{1}{R} \sum_{r = 1}^R p \left( \boldsymbol y_{t_{n}}^{i} \mid \theta_l^{\left( r \right)}, \mathcal M_l \right) \right) -  \ln \left( \frac{1}{R} \sum_{r = 1}^R p \left( \boldsymbol y_{t_{n}}^{i} \mid \theta_{l^\prime}^{\left( r \right)}, \mathcal M_{l^\prime} \right) \right) \right]
\end{equation}
where a positive value provides support for $\mathcal M_{l}$ over $\mathcal M_{l^\prime}$ and vice versa \citep{vehtari2017practical}.
The advantage of this scoring rule is that it considers the number and timing of points jointly when assessing the fit of each model to $\boldsymbol y_{\ts}$.

Our second approach considers a more common prediction problem in the context of online discussion trees: can we predict the eventual size of a discussion?
Formally, can we predict $n_i$ at time $a_i$, given that we have observed $\boldsymbol y_{t_{n}}^{i}$ up to time $t_{i 1} + s < a_i$, where $s > 0$ is the \textit{learning time interval}. 
As described in Section \ref{sec:simulating}, our generative model allows us to simulate a cluster over the interval $\left[t_{i 1} + s, a_i \right)$ given its observed trajectory over $\left[t_{i1}, t_{i 1} + s \right)$.
Thus, predicting the cluster size $n_i$ with $\mathcal M_l$ is relatively straightforward. 
For each $\theta_l^{\left( r \right)}$, we propagate the observation over $\left[t_{i1}, t_{i 1} + s \right)$ up to time $a_i$ and count the number of nodes in the simulated cluster, generating the predictions $n_{li}^{\left( 1 \right)}, \dots, n_{li}^{\left( R \right)}$.
We assess the quality of these predictions via an unbiased estimator for the continuous ranked probability score ($\crps$) \citep{zamo2018estimation}. As a first step, we sort $n_{li}^{\left( 1 \right)}, \dots, n_{li}^{\left( R \right)}$ into ascending order.
We then compute
\begin{equation}
    \widehat{\crps}_l  =  \frac{1}{S^\prime} \sum_{i = 1}^{S^\prime} \left( \frac{1}{R} \sum_{r = 1}^R \left| n_{li}^{\left( r \right)} - n_{i} \right| + \widehat{\phi}_{li0} - 2 \widehat{\phi}_{li1} \right)
\end{equation}
where $\widehat{\phi}_{li0} = \frac{1}{R} \sum_{r = 1}^{R} n_{li}^{\left( r \right)}$ and $\widehat{\phi}_{li1} = \frac{1}{R \left( R-1 \right)} \sum_{r = 1}^R \left( r - 1 \right) n_{li}^{\left( r \right)}$. 
This computation provides a proper scoring rule for our predictions where smaller values of $\widehat{\crps}_l$ indicate better predictive performance.
Finally, we compare $\mathcal M_l$ to a baseline via the continuous ranked probability skill score ($\crpss$), which we estimate as
\begin{equation}
    \widehat{\operatorname{crpss}_l} = 1 - \frac{\widehat{\crps}_l}{\widehat{\crps}_0}
\end{equation}
where we compute $\widehat{\crps}_0$ from the empirical cdf for cluster sizes implied by $\boldsymbol y_{\tr}$.
Interpreting the $\crpss$ is straightforward. 
Positive values imply that $\mathcal M_l$ improves on the baseline, with 1 representing perfect predictions, while negative values imply that the model underperforms relative to the baseline.

\section{Analysis and results}
\label{sec:results}

\subsection{Data description}
\label{sec:data_description}

As described in Section \ref{sec:reddit}, we analyse discussions on the \texttt{r/ireland} subreddit from April 1 until May 12, 2019.
This data set, which we denote $\boldsymbol y$, consists of 117,787 nodes over 38,616 discussions, where each node is associated with a unique identifier, a timestamp, and, where applicable, the identifier of its parent node. 
We treat each discussion as an independent cluster $\boldsymbol y_{t_{n}}^{i}$, where times $\boldsymbol t_i$ represent the number of hours since 00:00:00 on April 1, 2019.

Our Bayesian inference scheme is computationally expensive despite the $\mathcal O \left( n \right)$ cost associated with our generative model likelihood.
Thus, in the interests of reproducibility, we restrict our analyses to subsets of $\boldsymbol y$.
These subsets are nevertheless sufficient to draw substantive conclusions from the data.
Our first restriction is based on the observation that we only ever observe a discussion up to some point in time, and, in general, there is no guarantee that further comments do not occur after this point.
As such, for the cluster $\boldsymbol y_{t_{n}}^{i}$, we only `observe' the 48 hours following $t_{i1}$ such that $t_{ij} \in \left[ t_{i1}, a_i \right)$ and $a_i = t_{i1} + 48$ for all $i, j$.
Given that approximately $98.5\%$ of all discussions in $\boldsymbol y$ have no comments recorded more than 48 hours after the post is submitted, this seems a reasonable interval to consider.

Our training data consists of a random sample of $S = 2,017$ discussions seeded during the first three weeks of our analysis period, from Monday, April 1 to Sunday, April 21 inclusive.
As such, $\boldsymbol y_{\tr}$ has $N = \sum_{i = 1}^S n_i = 5,891$ training points in total and includes $10\%$ of the posts submitted in the first three weeks of the observation interval.

Our testing data is a random sample of $S^\prime = 3,716$ discussions arising from posts submitted between Monday, April 1 and Friday, May 10 inclusive. 
This allows us to observe the first 48 hours of each discussion in the testing set.
In this case, we have $N^\prime = \sum_{i = 1}^{S^\prime} n_i = 11,321$ testing points in total and $\boldsymbol y_{\ts}$ consists of $10\%$ of all posts submitted in the observation interval for which we observe at least 48 hours discussion.
No clusters in $\boldsymbol y_{\tr}$ are included within $\boldsymbol y_{\ts}$. 

\subsection{Candidate Models}

Our analysis considers a sequence of candidate models of increasing complexity, starting with an `off-the-shelf' Hawkes process and concluding with the generative model described in Section \ref{sec:generative_model}.

Our first model, denoted $\mathcal M_1$, is a Hawkes process with homogeneous offspring processes for each point, regardless of immigration status or arrival time.
That is, $\mathcal M_1$ omits the heterogeneous individual reproduction numbers, circadian rhythm, and the distinction between offspring processes for immigrants and offspring described in our generative model.
We encode these assumptions by assuming the sinusoidal basis frequencies $\boldsymbol \omega = \emptyset$ and dispersion parameters $\psi_1, \psi_2 \to \infty$.
The decay rates are then $\boldsymbol \eta = \eta_1 \mathbf 1_2$, and reproduction numbers are $\boldsymbol \mu = \mu_1 \mathbf 1_2$.
In effect, $\mathcal M_1$ is parameterised by $\theta_1 = \left\{\eta_1, \mu_1 \right\}$ such that $\nu_{ij} = \mu_1$ and $\gamma_{ij} \left( t \right) = \mu_1 \e \left( t - t_{ij} \mid \eta_1 \right)$ for all $j = 1, \dots, n_i$ and $i = 1, \dots, S$.
The inference scheme outlined in Section \ref{sec:parameter_inference} adapts to this setting easily.

The second model, $\mathcal M_2$, relaxes the strict homogeneity assumption in $\mathcal M_1$ by introducing different offspring processes for immigrants and offspring. Under $\mathcal M_2$, we have decay rates $\boldsymbol \eta = \left( \eta_1, \eta_2 \right)$ and reproduction numbers $\boldsymbol \mu = \left(\mu_1, \mu_2 \right)^\prime$ such that $\theta_2 = \left\{\eta_1, \eta_2, \mu_1, \mu_2 \right\}$. This model allows us to test whether the data support our assumption that the offspring process differs for immigrant and offspring points. 

Our third model, $\mathcal M_3$, extends  $\mathcal M_2$ by introducing the activity function $\alpha \left( t \right)$ via $K = 2$ sinusoidal basis frequencies $\boldsymbol \omega = 2 \pi \boldsymbol f$ where $\boldsymbol f = \frac{1}{24} \left( 1, 2 \right)^\prime$. That is to say, we choose two sinusoidal basis frequencies corresponding to one and two oscillations per day, respectively. We present an analysis supporting this choice in Appendix \ref{app:circadian}. Within this model, we have $\gamma_{i1} \left( t \right) = \mu_1 \, \alpha \left( t \right) \e \left( t - t_{ij} \mid \eta_1 \right)$ and $\gamma_{i1} \left( t \right) = \mu_2 \, \alpha \left( t \right) \e \left( t - t_{ij} \mid \eta_2 \right)$ for all $j = 2, \dots, n_i$ and $i = 1, \dots, S$. 
Thus, we have $\theta_3 = \left\{ \alpha_{1}, \dots, \alpha_{2K}, \eta_1, \eta_2, \mu_1, \mu_2 \right\}$.

The most complex model considered, $\mathcal{M}_4$, is the generative model described in Section \ref{sec:generative_model}, complete with dispersion parameters $\boldsymbol \psi = \left(\psi_1, \psi_2 \right)^\prime$. 
This model allows for over-dispersed offspring distributions via heterogeneous individual reproduction numbers and is parameterised by $\theta_4 = \left\{ \alpha_{1}, \dots, \alpha_{2K}, \eta_1, \eta_2, \mu_1, \mu_2, \psi_1, \psi_2 \right\}$.

Finally, we include one additional model, $\mathcal M_5$, where only immigrant points have heterogeneous individual reproduction numbers. That is to say, we include the dispersion parameter for the immigrant offspring process $\psi_1$, but assume that $\psi_2 \to \infty$.
The inclusion of this model, for which we have parameters $\theta_5 = \left\{ \alpha_{1}, \dots, \alpha_{2K}, \eta_1, \eta_2, \mu_1, \mu_2, \psi_1 \right\}$, allows us to assess whether our model for heterogeneous offspring processes is suitable for both immigrants and offspring.
With that, we have specified the full set of candidate models, summarised in Table \ref{tab:candidate_models}.

\begin{table}[ht]
    \centering
    \begin{tabular}{ |c|p{0.2\textwidth}|p{0.2\textwidth}|p{0.2\textwidth}|p{0.2\textwidth}| } 
         \hline
          & $\boldsymbol \mu$ & $\boldsymbol \psi$ & $\boldsymbol \eta$ & $\boldsymbol \omega$  \\ 
         \hline
         $\mathcal M_1$ 
         & $ \mu_1$ & $\psi_1, \psi_2 \to \infty$ & $ \eta_1$ & $\emptyset$ \\ 
         \hline
         $\mathcal M_2$ 
         & $\left(\mu_1, \mu_2 \right)^\prime$ & $\psi_1, \psi_2 \to \infty$ & $\left(\eta_1, \eta_2 \right)^\prime$ & $\emptyset$ \\ 
         \hline
         $\mathcal M_3$ 
         & $\left(\mu_1, \mu_2 \right)^\prime$ & $\psi_1, \psi_2 \to \infty$ & $\left(\eta_1, \eta_2 \right)^\prime$ & $\left( 2 \pi / 24 \right) * \left( 1, 2, 3 \right)^\prime$ \\ 
         \hline
         $\mathcal M_4$ 
         & $\left(\mu_1, \mu_2 \right)^\prime$ & $\left(\psi_1, \psi_2 \right)^\prime$ & $\left(\eta_1, \eta_2 \right)^\prime$ & $\left( 2 \pi / 24 \right) * \left( 1, 2, 3 \right)^\prime$ \\ 
         \hline
         $\mathcal M_5$ 
         & $\left(\mu_1, \mu_2 \right)^\prime$ & $\psi_1 > 0, \psi_2 \to \infty$ & $\left(\eta_1, \eta_2 \right)^\prime$ & $\left( 2 \pi / 24 \right) * \left( 1, 2, 3 \right)^\prime$ \\ 
         \hline
        \end{tabular}
    \caption{A summary of the parameters in each candidate model for discussion trees on the \texttt{r/ireland} subreddit. Unless otherwise stated, the domain for each parameter is the same as for the generative model presented in Section \ref{sec:generative_model}.}
    \label{tab:candidate_models}
\end{table}

\subsection{Inference}
\label{sec:inference}

The inference scheme described in Section \ref{sec:parameter_inference} allows us to fit each of the candidate models to $\boldsymbol{y}_{\tr}$ by sampling $p \left( \theta_l \mid \boldsymbol{y}_{\tr}, \mathcal M_l \right)$ for $l = 1. \dots, 5$.
For each of the candidate models, we sample 4 Markov chains of length 1,000 from their respective posterior distributions after a warm-up of 1,000 samples. 
In each case, these Markov chains satisfy standard convergence criteria with $\hat R \leq 1.01$ for all parameters \citep{vehtari2021rank}.

\begin{figure}[ht]
    \centering
    \includegraphics[width=0.9\textwidth]{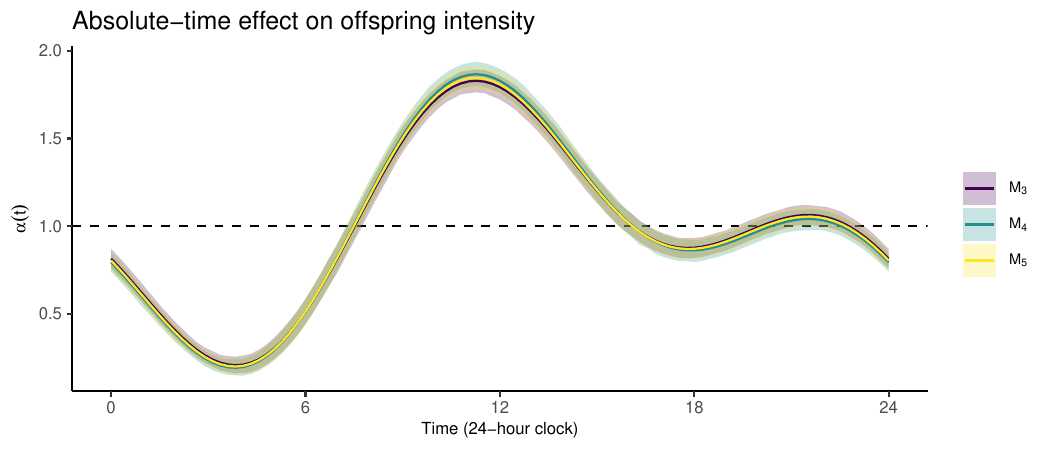}
    \caption{The mean and 95\% posterior probability intervals for $\alpha \left( t \right)$ over its 24-hour period under $\mathcal{M}_3$, $\mathcal{M}_4$, and $\mathcal{M}_5$. The dashed horizontal line at 1 represents the case where no absolute-time component is considered and is the implicit value of $\alpha \left( t \right)$ under $\mathcal M_1$ and $\mathcal M_2$.
    We find close agreement between all 3 models. In each case, the offspring intensity is inflated from approx 08:00 to 16:00, indicating that discussions on the \texttt{r/ireland} subreddit are most active during this time. Activity levels off from 16:00 to 22:00, before reducing overnight. Note that all times are reported in Irish Summer Time (IST). This suggests that users hoping to spark a lively discussion within this subreddit should submit posts in the morning to take advantage of this elevated offspring intensity.}
    \label{fig:candidate_activity}
\end{figure}

We present summaries of the marginal posteriors for parameters $\boldsymbol \eta$, $\boldsymbol \mu$, and $\boldsymbol \psi$ in each model in Table \ref{tab:parameter_estimates}.
To summarise the marginal posterior over $\boldsymbol \alpha$, we present $\alpha \left( t \right)$ for $t \in \left[0, 24 \right]$ in Figure \ref{fig:candidate_activity}, covering the full period of 24 hours.

Focusing on $\mathcal{M}_4$, the generative model proposed in Section \ref{sec:generative_model}, our analysis finds that the offspring processes for immigrants and offspring are indeed different.
While the reproduction numbers $\mu_1$ and $\mu_2$ are broadly similar, $\mu_1 \in \left(0.61, 0.7 \right)$ and $\mu_2 \in \left(0.62, 0.68 \right)$ with 95\% posterior probability, the decay rates $\eta_1$ and $\eta_2$ do not overlap, such that $\eta_1 \in \left(0.24, 0.26 \right)$ and $\eta_2 \in \left(0.33, 0.36 \right)$ with 95\% posterior probability.
This implies that when we average over $\alpha \left( t \right)$, the expected generation interval from immigrants is 3.8--4.2 hours, while it is 2.8--3.1 hours for offspring.
In addition, we find that $\psi_1 \in \left( 0.91, 1.38 \right)$ and $\psi_2 \in \left(4.30, 10.04 \right)$ with 95\% posterior probability. Following Section~\ref{sec:epidemic_dynamics}, we infer, for example, that the 20\% most infectious immigrants give rise to 47--53\% of expected offspring while 58--62\% generate none. Similarly, the $20\%$ most infectious offspring give rise to 29--34\% of expected offspring, and 52-55\% generate none.
Thus, individual reproduction numbers for immigrant points are moderately heterogeneous while those for offspring are relatively homogeneous, suggesting that homogeneous offspring processes may be suitable for this data.

Finally, our inference on $\alpha \left( t \right)$ is in agreement across $\mathcal M_3$, $\mathcal M_4$, and $\mathcal M_5$.
We find that $\alpha \left( \cdot \right)$ amplifies the offspring intensity functions between 09:00 and 16:00 each day, peaking between 10:00--12:00, and then levels off until approximately 22:00.
The activity function dampens offspring intensity functions overnight, reaching a trough between 04:00--06:00.
This suggests that posts submitted in the morning are more likely to generate large discussion trees than those made late at night due to higher overall activity levels when they are most likely to generate comments.
Section \ref{sec:predictive_performance} examines this intuition more thoroughly.

\begin{table}[ht]
\centering
\begin{tabular}{|c|c|c|c|c|c|c|}
  \hline
Model & $\mu_1$ & $\mu_2$ & $\eta_1$ & $\eta_2$ & $\psi_1$ & $\psi_2$ \\ 
  \hline
$\mathcal M_1$ & 0.66 (0.01) & - & 0.33 (0.01) & - & - & - \\ 
  $\mathcal M_2$ & 0.65 (0.02) & 0.67 (0.01) & 0.27 (0.01) & 0.38 (0.01) & - & - \\ 
  $\mathcal M_3$ & 0.64 (0.02) & 0.64 (0.01) & 0.25 (0.01) & 0.34 (0.01) & - & - \\ 
  $\mathcal M_4$ & 0.65 (0.02) & 0.65 (0.01) & 0.25 (0.01) & 0.34 (0.01) & 1.15 (0.12) & 6.99 (1.58) \\ 
  $\mathcal M_5$ & 0.65 (0.02) & 0.64 (0.01) & 0.25 (0.01) & 0.34 (0.01) & 1.15 (0.12) & - \\ 
   \hline
\end{tabular}
\caption{Posterior mean (standard deviation) for the parameters $\boldsymbol \eta$, $\boldsymbol \mu$, and $\boldsymbol \psi$ within each of our candidate models. The broad agreement on $\boldsymbol \mu$ across all the candidate models suggests that the expected number of offspring does not differ between immigrants and offspring. Differences in the offspring distributions for immigrants and offspring are manifest in the memory decay rate, which indicates that the expected generation interval for immigrants is longer than that for offspring. Finally, the values for $\boldsymbol \psi$ inferred by $\mathcal M_4$ suggest that immigrant points have a moderately heterogeneous offspring process while the offspring process for offspring is relatively homogeneous. As such, we include $\mathcal M_5$ in our analysis, which assumes heterogeneous immigrant and homogeneous offspring processes, respectively.} 
\label{tab:parameter_estimates}
\end{table}

While our examination of the posterior parameter distributions suggests that models $\mathcal M_4$ or $\mathcal M_5$ are likely to offer the best fit to the training data, we formalise this analysis by estimating the evidence of each model and computing Bayes factors comparing each model to $\mathcal M_4$. 
Our results are presented in Table \ref{tab:bf_estimates}.
We find overwhelming evidence to support the inclusion of the circadian rhythm $\alpha \left( \cdot \right)$ in our models for empirical data.
This tells us that empirical offspring processes are time-dependent and do not follow a standard Hawkes process.
We also find decisive evidence, as defined by \cite{kass1995bayes}, supporting heterogeneous individual reproduction numbers for both immigrant and offspring points, although the evidence is strongest for immigrant points.
That is to say, $\mathcal M_4$ and $\mathcal M_5$ offer the best fit to the data, with the model evidence providing decisive support for $\mathcal M_4$.

\begin{table}[ht]
\centering
\begin{tabular}{|l|c|c|c|c|c|}
  \hline
 & $\mathcal {M}_1$ & $\mathcal {M}_2$ & $\mathcal {M}_3$ & $\mathcal {M}_4$ & $\mathcal {M}_5$ \\ 
  \hline
$\ln \mathcal{BF}_{l4}$ & -494.74 & -448.37 & -117.35 & 0.00 & -10.65 \\ 
   \hline
\end{tabular}
\caption{Estimated log Bayes factor for each candidate model relative to $\mathcal M_4$, that is, $\ln \mathcal{BF}_{l4}$ for $l = 1, \dots, 5$. Note that the evidence supporting each candidate model is estimated from the sampled posterior distribution $p \left( \theta \mid \boldsymbol y_{\operatorname{train}}, \mathcal M_l \right)$ via bridge sampling. In each case, the \texttt{bridgesampling} algorithm reports a coefficient of variation for the evidence estimate of $<0.005$, indicating that we have a precise estimate for each model evidence and, as a result, the corresponding Bayes factors. We find decisive evidence to support our inclusion of a circadian rhythm in the offspring intensity. Furthermore, we find decisive support for including heterogeneous immigrant and offspring reproduction numbers.} 
\label{tab:bf_estimates}
\end{table}

\subsection{Assessing predictive performance}
\label{sec:predictive_performance}

Given that the above analysis supports the inclusion of a circadian rhythm and heterogeneous individual reproduction numbers when modelling discussions on the \texttt{r/ireland} subreddit, we now examine how these conclusions apply to our training data set.
In addition, we examine whether these findings result in improved predictive models for discussions.
First, we compare the expected log cluster-wise predictive density on $\boldsymbol y_{\ts}$ for each model to that of the generative model $\mathcal M_4$. 
That is, we compute $\Delta \widehat{\lpd}_{l4}$ given $R = 100$ samples from $p \left( \theta \mid \boldsymbol y_{\tr}, \mathcal M_l \right)$ for $l = 1, \dots, 5$.

In the second experiment, we assess predictions for the discussion size at some point in the future, given that we observe some initial learning interval $s$.
In other words, how many points will be in a cluster at time $a$ given that we observe the cluster over $\left[t_1, t_1 + s \right)$  for $t_1 + s < a$. In this case, $a = t_1 + 48$ hours for each cluster.
We evaluate the predictive performance of each model in terms of $\widehat{\crpss}$ on $\boldsymbol y_{\ts}$ given $R = 100$ predictions for the sequence of learning intervals $s \in \left\{0, 2^0, 2^1, 2^2, 2^3 \right\}$.  Note that the maximum learning interval considered in this experiment is $2^3 = 8$ hours. This is is because $\widehat{\crpss} \to 1$ as $t_1 + s \to a$ for all of our candidate models, making it impossible to distinguish between those that skillfully predict the final cluster size and those that do not. In other words, it does not require much skill to predict the trajectory of a discussion when you have already observed all the comments that are likely to occur.

\begin{figure}[ht]
    \centering
    \includegraphics[width=0.9\textwidth]{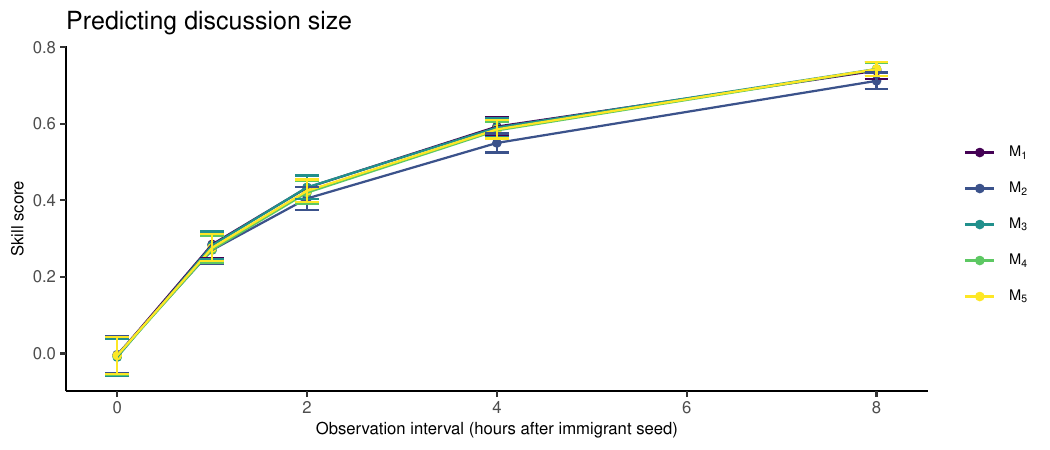}
    \caption{Continuous ranked probability skill scores for probabilistic predictions of the discussion size 48 hours after the post is seeded. The reference prediction is the empirical distribution of discussion sizes in $\boldsymbol y_{\tr}$. Error bars are estimated from the standard error of $\widehat{\crps}_l$ for $l = 1, \dots, 5$. Based on this analysis, we conclude that models that provide a superior fit to the data do not necessarily offer more skilled predictions for discussion size. In addition, none of the candidate models allows skilful \textit{ex-ante} prediction.}
    \label{fig:skill_score}
\end{figure}

This experimental set-up allows us to simultaneously compare each model and assess the impact of the learning interval on predictions.
Our analysis of $\widehat{\lpd}$ is presented in Table \ref{tab:lpd}, while $\widehat{\crpss}$ is illustrated in Figure \ref{fig:skill_score}.

\begin{table}[ht]
\centering
\begin{tabular}{|c|c|c|c|c|c|}
  \hline
 & $\mathcal {M}_1$ & $\mathcal {M}_2$ & $\mathcal {M}_3$ & $\mathcal {M}_4$ & $\mathcal {M}_5$ \\ 
  \hline
$\Delta \widehat{\operatorname{lpd}}_{l4}$ & -1004.6 (78.7) & -892.4 (71.8) & -177.5 (37.5) & 0.0 (0.0) & 17.2 (11.7) \\ 
   \hline
\end{tabular}
\caption{The difference (standard error) in log cluster-wise predictive density on $\boldsymbol y_{\operatorname{test}}$ between each model and $\mathcal M_4$. We find that $\mathcal M_4$ and $\mathcal M_5$ offer the best out-of-sample predictive performance in terms of $\operatorname{lpd}$, with $\mathcal M_5$ outperforming $\mathcal M_4$ slightly. This provides decisive support for the inclusion of a circadian rhythm and heterogeneous immigrant reproduction numbers in our model for online discussion on the \texttt{r/ireland} subreddit.} 
\label{tab:lpd}
\end{table}

We find that $\mathcal M_4$ and $\mathcal M_5$ provide the best fit to $\boldsymbol y_{\ts}$, each offering similar performance in terms of $\widehat{\lpd}$.
Both models comfortably outperform $\mathcal M_3$, which in turn is a significant improvement over either $\mathcal M_1$ or $\mathcal M_2$.
Thus, we find that modelling the circadian rhythm and allowing for heterogeneous immigrant reproduction numbers in the \texttt{r/ireland} subreddit improves predictive models for discussions within this community.

While the analysis in Table \ref{tab:lpd} makes it clear that our generative model offers improved $\widehat{\lpd}$ performance over simpler Hawkes process models for online discussions, this does not necessarily mean that it will also improve predictions for the discussion size at future time points.
The becomes clear when we examine Figure \ref{fig:skill_score}, which presents $\widehat{\crpss}_l$ for $l = 1, \dots, 5$ at each value for $s$.

Our analysis shows that none of the candidate models offers \textit{ex-ante} prediction that beats the empirical baseline, which is to say that none of the candidate models accurately forecast the size of a discussion before the initial post is made.
Once we observe the learning interval, however, the skill of our predictive models improves rapidly, with an interval of between 2 and 4 hours sufficient to ensure each model has a skill score $> 0.5$.
Interestingly, there is no significant difference in skill scores between any candidate models.
This indicates that a simple Hawkes process model is as effective for predicting discussion size as the proposed generative model, at least when we wish to make predictions for the discussion size after 48 hours.
This leads us to conclude that modelling the within-community circadian rhythm and heterogeneous individual reproduction numbers do not necessarily improve predictions for a discussions size,
despite decisive evidence supporting the inclusion of a within-community circadian rhythm and heterogeneous immigrant reproduction numbers within our models. 

\subsection{Assessing goodness-of-fit}
\label{sec:gof}

\begin{figure}[ht]
    \centering
    \includegraphics[width=0.9\textwidth]{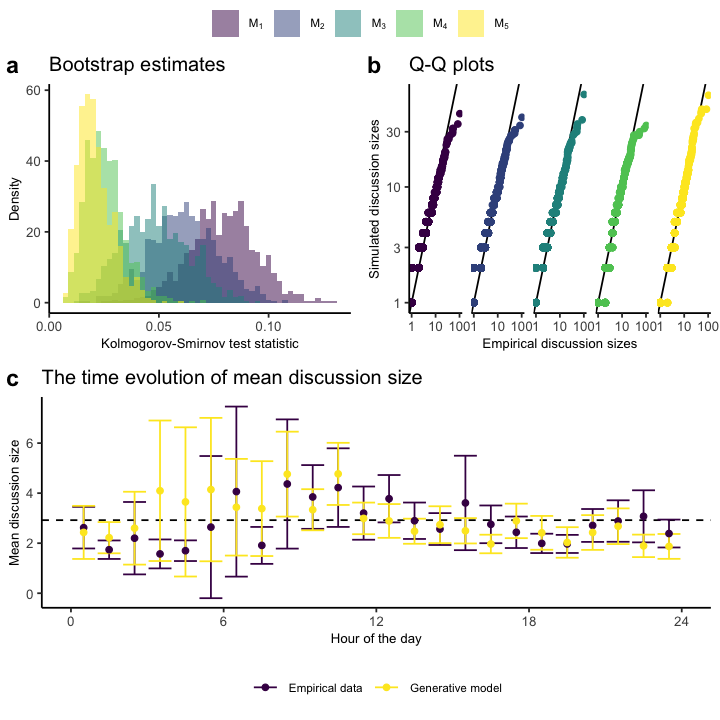}
    \caption{Assessing the goodness-of-fit of generative models for online discussion. Plot \textbf{a} presents bootstrap estimates for the Kolmogorov-Smirnov test statistic comparing the distribution of discussion sizes under each model to the empirical data and suggests that models $\mathcal{M}_4$ and $\mathcal{M}_5$ offer the best fit.
    Plot \textbf{b} presents Quantile-Quantile (Q-Q) plots of the empirical and simulated data, demonstrating that empirical data retains a long tail of large discussions relative to each simulated data set.
    Plot \textbf{c} illustrates the dependence of mean discussion size on the time users submit posts to the subreddit for both the empirical data and the data simulated from $\mathcal{M}_4$, the generative model. Points represent the mean discussion size for posts submitted at each hour of the day, error bars represent an interval of two times the standard error above and below the mean, and the dashed line represents the overall mean discussion size for $\boldsymbol y_{\tr}$.}
    \label{fig:gof}
\end{figure}

The final section of our analysis considers a goodness-of-fit analysis for each candidate model to the training data $\boldsymbol y_{\tr}$ based on an analysis of discussion sizes predicted from each model. To do this, we simulate data from each of the candidate models.
Given a sample from the posterior $p \left( \theta_l \mid \boldsymbol y_{\tr}, \mathcal M_l \right)$, we simulate a discussion for each immigrant point in $\boldsymbol y_{\tr}$ and note the size of the discussion. 
This process provides a simulated data set $\boldsymbol y_l^*$ of 2,017 discussions from $\mathcal M_l$ for $l = 1, \dots, 5$, where discussions share a common set of immigrant points.
Comparing these simulated discussions to the empirical data allows us to assess the goodness-of-fit for each model.
Our analysis is presented in Figure \ref{fig:gof}.

Our first assessment, presented in Figure \ref{fig:gof}\textbf{a}, considers the two-sample Kolmogorov-Smirnov (KS) test statistic comparing the distribution of discussion sizes in $\boldsymbol y_l^*$ to that of $\boldsymbol y_{\tr}$.
To quantify uncertainty on this test statistic, we construct bootstrap estimates for our test statistic via 1,000 bootstrap samples.
Each sample consists of the discussions in $\boldsymbol y_{\tr}$ and $\boldsymbol y_l^*$ associated with 2,017 immigrant points sampled with replacement.
Note that the same 2,017 immigrant points are used to construct KS test statistics for all five models within each bootstrap sample.
We find that $\mathcal M_5$, the generative model allowing for heterogeneous immigrant reproduction numbers, offers a reasonable fit to empirical data. Bootstrapped estimates for the KS test statistic are less than those for $\mathcal M_1$, $\mathcal M_2$, or $\mathcal M_3$ in 99.7\%, 97.7\% and 93\% of bootstrapped samples, respectively. It is more difficult to distinguish between the two candidate models, $\mathcal M_4$ and $\mathcal M_5$, that allow for heterogeneous individual reproduction numbers as the KS statistic for $\mathcal M_5$ is less than that for $\mathcal M_4$ in only 65\% of bootstrapped samples. In summary, we find that the distribution of discussion sizes under $\mathcal M_4$ and $\mathcal M_5$ is closest to that of empirical data.

In Figure \ref{fig:gof}\textbf{b}, we examine the Q-Q plot comparing the distribution of discussion sizes for $\boldsymbol y_{\tr}$ to that of $\boldsymbol y_l^*$ for $l = 1, \dots, 5$.
We see that each of the Q-Q plots is broadly similar. 
The quantiles of simulated data begin to depart from the empirical data for discussions of around thirty or more nodes, implying that empirical data has a longer tail of large discussions than any of our simulated data sets, although the sample simulated from $\mathcal M_5$ appears to come closest.

Finally, in Figure \ref{fig:gof}\textbf{c}, we present the mean discussion size arising from posts submitted in each hour of the day for both the empirical data and the data simulated from $\mathcal{M}_4$, the generative model.
We see that the mean discussion size varies throughout the day.
For empirical data, this mean is largest between 07:00 and 13:00, whilst in the simulated data, it is elevated between 04:00 and 11:00 before settling below 3 from 14:00 to 02:00.
Thus, we conclude that the model can reproduce variation in the data due to the circadian rhythm of the online community hosted by \texttt{r/ireland}.


\section{Discussion}
\label{sec:discussion}

In this report, we have developed a novel model for online discussion trees and applied it to data taken from the \texttt{r/ireland} subreddit.
Our goal with this model was to capture the over-dispersed reply distribution and circadian rhythm typically associated with empirical data of this nature.
To this end, we drew parallels between online discussion and epidemic disease, where over-dispersed offspring distributions manifest as superspreading dynamics.
As such, we introduced latent individual reproduction numbers modelling variation in the expected number of replies to each node in the discussion.
Within this framework, more heterogeneous individual reproduction numbers imply a more over-dispersed reply distribution.
In addition, we introduced a sinusoidal activity function modelling the circadian rhythm of the \texttt{r/ireland} community by modulating offspring intensities throughout the day.
Our analysis considered a set of candidate models ranging from a standard Hawkes process to the full generative model.
We fit these models to a subsample of our data by sampling the posterior with Stan. 
We then assessed each model via estimates for the model evidence and out-of-sample predictive performance.
Finally, we examined the goodness-of-fit for each model by simulating discussions seeded by each post in our empirical data.

We draw several conclusions from this analysis.
Firstly, our proposed model describes the observed data well. 
Estimates for the evidence provide overwhelming support for modelling the circadian rhythm and including heterogeneous individual reproduction numbers for each post.
Such models delivered the best predictive performance, as measured by the log cluster-wise predictive density of each candidate model on our test set, and data simulated from this model has the closest fit to empirical data.
Furthermore, modelling the circadian rhythm of activity on the \texttt{r/ireland} subreddit allows us to estimate that the mean discussion size is
approximately 4 for posts submitted between 04:00 and 12:00 in the morning, while this mean is approximately 2.5 between 15:00 and 02:00. 
We also find that 58-62\% of posts generate no further discussion, with 95\% posterior probability.
Thus, the proposed model offers a general approach to generating insight into online discussion dynamics.

Our second conclusion relates to the difficulty associated with forecasting features of individual discussions.
In particular, the challenges associated with predicting a discussion's eventual size.
We found no evidence to suggest that any candidate model could produce reliable \textit{ex-ante} forecasts for discussion size. Furthermore, while a learning interval of 1--2 hours improved these models' predictive skill significantly, there was no perceptible difference between any of the models applied to this task.
Thus, despite the excellent fit of our generative model to data, modelling circadian rhythms and over-dispersed reply distributions may not improve predictive performance.
However, such results are not unexpected.
Only 5\% of discussions in our data set consist of ten or more nodes. 
Indeed, only 0.3\% have at least 50 nodes, the lower bound for inclusion in the analysis presented by \cite{medvedev2019modelling}.
As such, large discussions are exceedingly rare, making accurate \textit{ex-ante} forecasting of these events difficult, if not impossible.
Thus, it is reasonable for analysts to focus on developing coherent explanations for empirical data, a task for which our proposed methodology is well suited.

Our third conclusion relates to the flexibility and broad applicability of Hawkes processes.
All candidate models in our analysis fit within the same generative model structure, making it relatively straightforward to define and compare interpretable models for our data.
Furthermore, while we developed a Bayesian approach to parameter inference in this instance, we can also easily sample from this generative model, making simulation-based inference possible.

There remain several avenues for future work building on this research.
Firstly, given some observed marks $\kappa_j \in \mathcal{K}$, our modelling framework could be adapted to test for associations between specific marks and individual reproduction numbers where $\mathbb E \left[ \nu_j \right] = f \left( \kappa_j \right)$.
Such a model would provide a rigorous approach to identifying viral content.

A more challenging task would be to address some of this model's shortcomings. 
We noted in Section \ref{sec:gof} that while our generative model offers the best fit to empirical data, it did not capture the long tail of very large discussions.
While our generative model describes posts that fail to generate any further discussion well, it may miss important features of the offspring distribution for comments. 
Figure \ref{fig:reply_exploration}\textbf{b} indicated that the reply distribution for comments had a heavy tail, but our modelling suggested that their individual reproduction numbers were relatively homogeneous.
This issue might arise when we see more comments with one reply than expected under a Negative Binomial distribution for replies.
Such a distribution is consistent with a process where we have two types of replies to a comment.
If the first type is a reply from the general population of users on the subreddit, then the other type comes from the specific user that submitted the parent node to the comment.
Thus, we have a scenario where two users have a back-and-forth conversation within a broader discussion.
Such a process does not fit within our generative model; however, modelling these discussion dynamics may be crucial to capturing the long tail in empirical data.

Finally, we note that this work had drawn on methods developed for understanding the spread of disease through a population.
While empirical data on epidemics tend to be sparse, given a small amount of good-quality contact tracing data, the methods developed here may be adapted to this context.
Such an analysis could provide insight into transmission dynamics, allowing policymakers to model the effect of control measures.
Thus, just as we have been motivated by epidemiological techniques, we hope this work will inform future research on the spread of infectious diseases.

\begin{acks}[Acknowledgments]
The authors would like to thank J. O'Brien, who provided a raw data set of discussions on the \texttt{r/ireland} subreddit.
\end{acks}
\begin{funding}

The Insight Centre for Data Analytics is supported by Science Foundation Ireland under Grant Number 12/RC/2289$\_$P2.

\end{funding}

\begin{supplement}
\stitle{Code supplement}
\sdescription{R code implementing the analysis presented here. Calls the R package which containing the required the data and algorithms. The code supplement is also available on Github at \href{https://github.com/jpmeagher/modelling_online_discussion_supplement}{github.com/jpmeagher/modelling\_online\_discussion\_supplement} while the R package is available at \href{https://github.com/jpmeagher/onlineMessageboardActivity}{github.com/jpmeagher/onlineMessageboardActivity}.}
\end{supplement}


\bibliographystyle{imsart-nameyear} 
\bibliography{bibliography}       


\begin{appendix}
\section{Evaluating the compensator}
\label{app:compensator_derivation}

To obtain equation (\ref{eq:compensator}) for the compensator of the ground process defined by (\ref{eq:conditional_intensity}) consider the following:
\begin{equation*}
    \begin{aligned}
        \Lambda^* \left( a \right) 
        &= \int_{t_{1}}^a \lambda^* \left( u \right) du, \\
        &= \int_{t_{1}}^a \boldsymbol \alpha^\prime \boldsymbol s_{\boldsymbol \omega} \left( u \right) \left( \nu_{1} \, \eta_1 \exp \left( - \eta_1 \left( u - t_{1} \right) \right) +  \sum_{j = 2}^{n_i} \nu_{j} \, \eta_2 \exp \left( - \eta_2 \left( u - t_{j} \right) \right) \right) du, \\
        &= \nu_{1} \, \int_{t_{1}}^a \boldsymbol \alpha^\prime \boldsymbol s_{\boldsymbol \omega} \left( u \right) \, \eta_1 \exp \left( - \eta_1 \left( u - t_{1} \right) \right)  du + \\
        & \hspace{1cm} \sum_{j = 2}^{n_i} \nu_{j} \int_{t_{j}}^a \boldsymbol \alpha^\prime \boldsymbol s_{\boldsymbol \omega} \left( u \right) \, \eta_2 \exp \left( - \eta_2 \left( u - t_{j} \right) \right) du , \\
    \end{aligned}
\end{equation*}
where the final line is a consequence of Fubini's theorem.
Now, because the portion of the compensator associated with the immigrant is of the same form as the portion associated with each offspring, we derive a general expression for the compensator of each point. As such, consider
\begin{equation*}
    \begin{aligned}
         & \nu \int_{t}^a \boldsymbol \alpha^\prime \boldsymbol s_{\boldsymbol \omega} \left( u \right)  \, \eta \exp \left( - \eta \left( u - t \right) \right) du \\
         &= \nu \int_{0}^{a - t} \boldsymbol \alpha^\prime \boldsymbol s_{\boldsymbol \omega} \left( u + t \right)  \, \eta \exp \left( - \eta u \right) du, \\
         &= \nu \int_{0}^{a - t} \sum_{k = 0}^{2K} \alpha_k \, s_k \left( u + t \right)  \, \eta \exp \left( - \eta u \right) du, \\
         &= \nu \sum_{k = 0}^{2K} \alpha_k \int_{0}^{a - t} w_k \left( u, t, \eta \right) du, \\
         &= \nu \sum_{k = 0}^{2K} \alpha_k W_k \left(t, a, \eta\right), \\
         &= \nu \, \boldsymbol \alpha^\prime \boldsymbol W \left(t, a, \eta\right), 
    \end{aligned}
\end{equation*}
where we have dropped unnecessary indexing variables for notational ease and defined $w_k \left( u, t, \eta \right) = s_k \left( u + t \right)  \, \eta \exp \left( - \eta u \right)$ for $k = 0, \dots, 2K$, which we refer to as an exponentially weighted sinusoidal basis function. Thus, in order to evaluate the compensator in (\ref{eq:compensator}), we need to compute $W_k \left(t, a, \eta \right)$, the integral of our exponentially weighted sinusoidal basis function, for each $k = 0, 1 \dots, 2K$. Analytic expressions for these integrals are given by $W_0 \left(t, a, \eta \right) = F_\eta \left( a - t \right)$ and
\begin{equation*}
    \begin{aligned}
         &W_{2k - 1} \left(t, a, \eta \right) \\
         &= \frac{\eta}{\eta^2 + \omega_k^2} \left( - \exp \left( \eta \left(t - a \right) \right) \left(
         \eta \sin \left( a \omega_k \right) + \omega_k \cos \left( a \omega_k \right)
         \right)
         + \eta \sin \left( t \omega_k \right) + \omega_k \cos \left( t \omega_k \right)
         \right), \\
         &W_{2k} \left(t, a, \eta \right) \\
         &= \frac{\eta}{\eta^2 + \omega_k^2}
         \left( \exp \left( \eta \left(t - a \right) \right) \left(
         \omega_k \sin \left( a \omega_k \right) - \eta \cos \left( a \omega_k \right)
         \right)
         + \eta \cos \left( t \omega_k \right) - \omega_k \sin \left( t \omega_k \right)
         \right), \\
    \end{aligned}
\end{equation*}
where $ F_\eta \left( x \right) = \int_0^x \e \left( u \mid \eta \right) du$ is the exponential cumulative distribution function.

\section{Integrating over latent marks}
\label{app:integration}

To integrate over the latent marks in the complete-data likelihood presented in equation (\ref{eq:complete_likelihood}), we first consider the integral over the latent mark $\nu_j$ for $j \in \left\{2, \dots, n \right\}$ such that
\begin{equation}
\begin{aligned}
    \int p \left(\boldsymbol y_{t_n}, \boldsymbol \nu \mid \theta \right) d \nu_j
    &= \int \left[ \prod_{k = 2}^n \lambda^* \left( t_k \right) \right] \exp \left( - \Lambda^* \left( a \right)\right)  \left[ \prod_{k = 1}^n f^* \left( \nu_k \mid \beta_k \right) \right] d \nu_j, \\
    &= K_{-j} \int \nu_j^{z_j} \exp \left( - \nu_j c_{t_j} \right) \gam \left( \nu_j \mid \psi_2, \psi_2 / \mu_1 \right) d \nu_j, \\
    &= K_{-j} \frac{\left(\psi_2 / \mu_2 \right)^{\psi_2}}{\Gamma \left(\psi_2 \right)} \int \nu_j^{z_j + \psi_2 - 1} \exp \left( - \nu_j \left( c_{t_j} + \psi_2 / \mu_2 \right) \right) d \nu_j, \\
    &= K_{-j} \frac{\Gamma \left(z_j + \psi_2 \right)}{\Gamma \left(\psi_2 \right)}  \left( \frac{\psi_2}{\mu_2 c_{t_j} + \psi_2} \right)^{\psi_2} \left( \frac{\mu_2}{\mu_2 c_{t_j} + \psi_2} \right)^{z_j}.
\end{aligned}
\end{equation}
Here, we have the portion of the complete-data likelihood that is independent of $\nu_j$
\begin{equation*}
    K_{-j} = \left[ \prod_{k \neq j} \lambda^* \left( t_k \right) \right] \exp \left( - \nu_{1} \, \boldsymbol \alpha^\prime \boldsymbol W \left(t_{1}, a, \eta_1 \right) + \sum_{k \neq j} \nu_{k} \, \boldsymbol \alpha^\prime \boldsymbol W \left(t_{k}, a, \eta_2 \right) \right)  \left[ \prod_{k \neq j}^n f^* \left( \nu_k \mid \beta_j \right) \right],
\end{equation*}
the observed offspring $z_j = \sum_{k = 1}^n \beta_k = j$, and the contribution to compensator due to $t_j$ is $c_{t_j}$. 
Thus, we see that the distribution of $\nu_j$ given $\boldsymbol y_{t_n}$ is independent of the reproduction number associated with any other point such that
\begin{equation}
    p \left( \nu_j \mid \boldsymbol y_{t_n} \right) = \gam \left( \nu_j \mid z_j + \psi_2, c_{t_j} + \psi_2 / \mu_2 \right).
\end{equation}
Repeating this analysis for each latent mark, we obtain the model likelihood presented in (\ref{eq:model_likelihood}).

\section{Simulation Algorithms}
\label{app:simulation_algorithm}

\RestyleAlgo{ruled}

Our generative model assumes the Poisson process $\Phi_j$ has an intensity function $\gamma_j \left( t \right)$ of the form $\nu_j \alpha \left( t \right) \rho \left( t - t_j \mid \beta_j \right)$ where $\alpha \left( t \right) = \boldsymbol \alpha^\prime \boldsymbol s_{\boldsymbol \omega} \left( t \right)$ and $\rho \left( t - t_j \mid \beta_j \right)$ is the probability density function of an exponential random variable with memory decay $\eta$ denoted $\e \left(t - t_j \mid \eta \right)$.
To generate realisations from this Poisson process, we first identify a dominating Poisson process $\tilde \Phi_j$ with intensity $\tilde \gamma_j \left( t \right) \geq \gamma_j \left( t \right)$ for all $t$. To this end, we note that there exists some constant $K_0$ such that $K_0 \geq \alpha \left( t \right)$ for all $t \in \mathbb R$. Therefore, we have
\begin{equation}
    \tilde \gamma_j \left( t \right) =   K_0 \, \nu_j \, \rho \left( t - t_j \mid \beta_j \right) \geq \gamma_j \left( t \right),
\end{equation}
as required.
If we let $\Phi_j^*$ denote a realisation of the point processes $\Phi_j$ on the interval $\left( a, b \right)$, our simulation algorithm is presented in Algorithm \ref{alg:offspring}.

\begin{algorithm}
\caption{Simulating the offspring process}
\label{alg:offspring}
\KwIn{$t_j$, $a$, $b$, $\boldsymbol \omega$, $\nu_j$, $\boldsymbol \alpha$, $\eta$, $K_0$}
Initialise $\Phi_j^* \gets \emptyset$\;
Draw $\tilde z_j \sim \pois \left( \nu_j K_0 \left( F_{\eta} \left(b - t_j \right) - F_{\eta} \left(a - t_j \right)\right) \right)$\;
\If{
$\tilde z_j > 0$
}{
\For{$k = 1, \dots, \tilde z_j$}{
Draw $\tau_k \sim \te \left( \eta, t_j - a, t_j - b \right)$ and $u_k \sim \unif \left(0, 1\right)$\;
Set $t_k \gets t_j + \tau_k$\;
\If{$u_k \leq \boldsymbol \alpha^\prime \boldsymbol s_{\boldsymbol \omega} \left( t_k \right) / K_0$}{
$\Phi_j^* \gets \Phi_j^* \cup t_k$\;
}
}
}
\Return{$\Phi_j^*$}
\end{algorithm}

In Algorithm \ref{alg:offspring} we have defined $\tilde z_j$, the total number of offspring from the process on the interval from the Poisson pmf where
\begin{equation*}
\begin{aligned}
    p \left( \tilde z_j \mid t_j, \nu_j, K_0, \eta \right) 
    &= \pois \left( \tilde z_j \mid  \int_{a}^{b} \tilde \gamma_j \left( u \right) du \right), \\
    &= \pois \left( \tilde z_j \mid \nu_j K_0 \left( F_{\eta} \left(b - t_j \right) - F_{\eta} \left(a - t_j \right)\right)\right),
\end{aligned}
\end{equation*}
and $F_{\eta} \left( x \right)$ is the exponential cdf.
We also have generation intervals $\tau_k$, which follow a truncated exponential distribution.
Finally, we note that for loops in this and the subsequent Algorithm may be vectorised for computational efficiency.
Given that we can now simulate realisations of the offspring process $\Phi_j$ on the interval $\left(a, b \right)$, our algorithm for simulating $\boldsymbol y_{n+m}^*$ is presented in Algorithm \ref{alg:cluster}.

\begin{algorithm}
\caption{Propagating a cluster}
\label{alg:cluster}
\KwIn{$\boldsymbol y_{t_n} = \left( \boldsymbol t, \boldsymbol \beta \right)$, $a$, $b$, $\boldsymbol \omega$, $\boldsymbol \mu$, $\boldsymbol \psi$, $\boldsymbol \alpha$, $\eta$, $K_0$}
Initialise $\boldsymbol z \gets \emptyset$\;
\For{$j = 1, \dots, n$}{
Set $z_j \gets \sum_{k = 1}^n \delta \left( \beta_k = j \right)$ and $\boldsymbol z \gets \boldsymbol z \cup z_j$\;
}
Draw $\nu_1 \sim \gam \left( z_1 + \psi_1, \boldsymbol \alpha^\prime \boldsymbol W \left(t_1, a, \eta_1 \right) + \frac{\psi_1}{\mu_1} \right)$ and set $\boldsymbol \nu^* \gets \nu_1$\;
\For{$j = 2, \dots, n$}{
Draw $\nu_j \sim \gam \left( z_j + \psi_2, \boldsymbol \alpha^\prime \boldsymbol W \left(t_j, a, \eta_2 \right) + \frac{\psi_2}{\mu_2} \right)$ and set $\boldsymbol \nu^* \gets \boldsymbol \nu^* \cup \nu_j$
}
Initialise $\boldsymbol t^* \gets \boldsymbol t$, $\boldsymbol \beta^* \gets \boldsymbol \beta$, and $m \gets 0$\;
Simulate $\Phi_1^*$ by Algorithm \ref{alg:offspring}\;
\If{$\Phi_1^* \neq \emptyset$}{
\For{$t_k \in \Phi_1^*$}{
Set $m \gets m+1$\;
Draw $\nu_k \sim \gam \left( \psi_2, \frac{\psi_2}{\mu_2} \right)$\;
Set $\boldsymbol t^* \gets \boldsymbol t^* \cup t_k$, $\boldsymbol \beta^* \gets \boldsymbol \beta^* \cup 1$, $\boldsymbol \nu^* \gets \boldsymbol \nu^* \cup \nu_k$\;
}
}
Set $j \gets 2$\;
\While{$j \leq n + m$}{
Simulate $\Phi_j^*$ by Algorithm \ref{alg:offspring}\;
\If{$\Phi_j^* \neq \emptyset$}{
\For{$t_k \in \Phi_j^*$}{
Set $m \gets m+1$\;
Draw $\nu_k \sim \gam \left( \psi_2, \frac{\psi_2}{\mu_2} \right)$\;
Set $\boldsymbol t^* \gets \boldsymbol t^* \cup t_k$, $\boldsymbol \beta^* \gets \boldsymbol \beta^* \cup j$, $\boldsymbol \nu^* \gets \boldsymbol \nu^* \cup \nu_k$\;
}
}
Set $j \gets j+1$\;
}
\Return{$\boldsymbol y_{t_{n+m}}^* = \left(\boldsymbol t^*, \boldsymbol \beta^* \right)$}
\end{algorithm}

\section{Modelling the circadian rhythm}
\label{app:circadian}

Our analysis makes the simplifying assumption that sinusoidal basis functions for $\alpha \left( t \right)$ may be treated as fixed hyper-parameters of our generative model.
This assumption is motivated by known patterns of human behaviour.
Human users of online boards tend to follow diurnal activity patterns and online communities are often based in some geographical location.
As such, it is reasonable to assume that online activity patterns follow a 24-hour cycle.
Thus, we do not include sinusoidal basis frequencies in the set of parameters inferred by sampling from the likelihood (\ref{eq:model_likelihood}), which is a complicated, non-linear function of $\boldsymbol \omega$.

\begin{figure}[ht]
    \centering
    \includegraphics[width=0.9\textwidth]{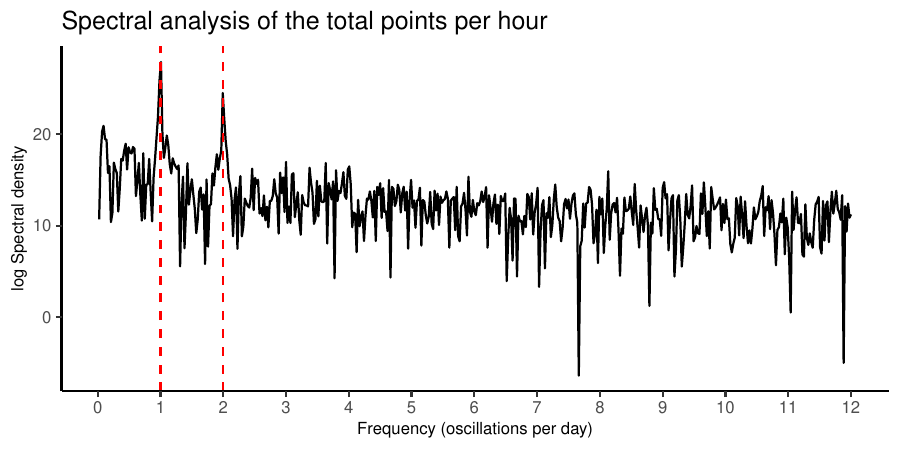}
    \caption{A periodogram estimating the power spectral density function for the point process of posts and comments on the \texttt{r/ireland} subreddit. The periodogram is dominated by two frequency components marked by the dashed vertical lines. These frequencies correspond to one and two oscillations per day, supporting our assumption that the intensity with which points arrive will follow a circadian rhythm and that frequency components for sinusoidal basis functions modelling this rhythm can be specified a priori.}
    \label{fig:spectral_analysis}
\end{figure}

That said, we performed some preliminary analysis to see if this pattern is borne out by the data.
Counting the number of points per hour for the full data set described in Section \ref{sec:data_description} provides a time series of hourly counts. 
The periodogram of this series, which estimates the power spectral density function for the point process \citep{bartlett1963spectral}, is presented in Figure \ref{fig:spectral_analysis}.
This analysis identifies two major peaks in the spectral density corresponding to one and two oscillations per day.
While this is far from a rigorous spectral analysis of the offspring process in our generative model, it suggests that our data follows a circadian rhythm dominated by two frequency components, offering comfort that our modelling assumptions are appropriate.

Confident that our data follows a circadian rhythm, we must select $K$, the number of sinusoidal basis functions for $\alpha \left( t \right)$.
Figure \ref{fig:spectral_analysis} suggests that $K = 2$ is most appropriate; however, this periodogram relates to the overall point process on the \texttt{r/ireland} subreddit rather than the offspring process in our generative model.
As such, it may be dominated by the exogenous process generating posts.
Therefore, we undertake further evidence-based analysis to ensure that this is a sensible choice for $K$.

For this analysis, let $\mathcal M_K$ denote the full generative model outlined in Section \ref{sec:generative_model} with heterogeneous individual reproduction numbers and $K$ sinusoidal basis frequencies. 
We consider each of the models defined by $K \in \left\{1, \dots, 4 \right\}$ where $\omega_k = 2 \pi \left( k / 24 \right)$ for $k = 1, \dots, K$. 
That is to say; we consider models specified by one, two, three, and four sinusoidal basis frequencies corresponding to one, two, three, and four oscillations per day, respectively.
We fit these models in Stan, sampling 4 Markov chains of length 1000 from the posterior distribution (\ref{eq:posterior_definition}) after a warm-up of 1000 samples and obtain the bridge sampling estimate of the evidence for each model.
We then compute Bayes factors for each model, treating $K = 2$ as our baseline model.
The results of this analysis are presented in Table \ref{tab:freq_bf_estimates}.

\begin{table}[ht]
\centering
\begin{tabular}{|l|c|c|c|c|}
  \hline
 & $\mathcal {M}_1$ & $\mathcal {M}_2$ & $\mathcal {M}_3$ & $\mathcal {M}_4$ \\ 
  \hline
$\ln \mathcal{BF}_{l2}$ & -170.23 & 0.00 & 5.01 & 16.12 \\ 
   \hline
\end{tabular}
\caption{Estimated log Bayes factor for each candidate model relative to $\mathcal M_2$, that is, $\ln \mathcal{BF}_{l2}$ for $l = 1, \dots, 4$. The evidence supporting each candidate model is estimated from the sampled posterior distribution $p \left( \theta \mid \boldsymbol y_{\operatorname{train}}, \mathcal M_l \right)$ via bridge sampling. In each case, the \texttt{bridgesampling} algorithm reports a coefficient of variation for the evidence estimate of $<0.005$, indicating that we have a precise estimate for each model evidence. This analysis presents overwhelming evidence to support a choice of $K > 1$.} 
\label{tab:freq_bf_estimates}
\end{table}

We find overwhelming evidence supporting $K > 1$. The largest improvement in model evidence occurs when adding the second sinusoidal basis function, but the evidence supporting the third and fourth basis functions would be considered decisive \citep{kass1995bayes}. Despite this evidence, we prefer a more parsimonious model for data and only include additional basis functions when they alter our interpretation of the circadian rhythm of the online community.

\begin{figure}[ht]
    \centering
    \includegraphics[width=0.9\textwidth]{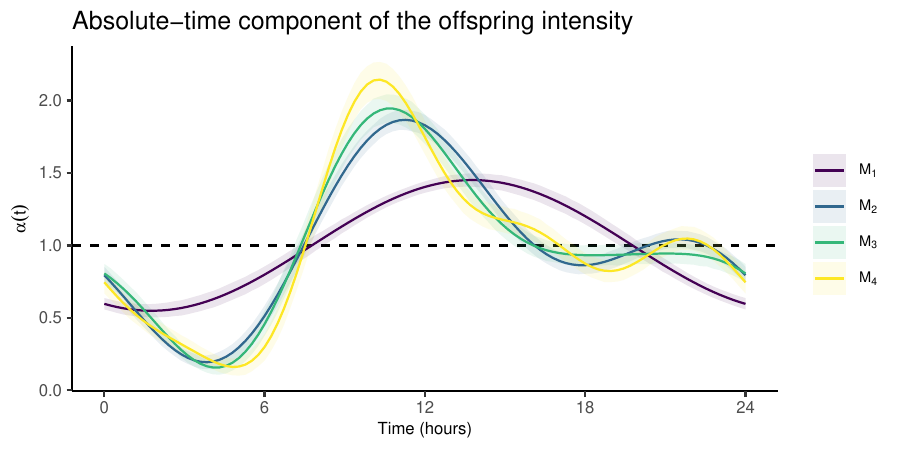}
    \caption{The absolute-time component of the offspring intensity function $\alpha \left( t \right)$ inferred using $K$ sinusoidal basis functions. Solid lines represent the posterior mean for each $\alpha \left( t \right)$ while shaded regions represent 95\% credible intervals. The horizontal dashed line highlights the relative deviance of each model from the baseline where $\alpha \left( t \right) = 1$ for all $t$. We find good agreement between $\mathcal M_2$ and $\mathcal M_3$, with consistent overlap of the respective credible intervals.}
    \label{fig:freq_activity_functions}
\end{figure}

Figure \ref{fig:freq_activity_functions} presents the function $\alpha \left( t \right)$ inferred by each model, showing broad agreement between models with $K \geq 2$, with each activity function identifying a peak just before noon and a trough at around 04:00.
Thus, we select $K = 2$ for analysis in the main text, which offers a robust model for our data that captures key feature of the \texttt{r/ireland} circadian rhythm.

\section{Modelling immigrant arrivals}

When the objective of our model for online board activity is to predict discussion size, then it is not necessary to specify the post-generating (immigrant) process.
If a more general model for online board activity was required, for example, if our objective was to predict or monitor overall activity on the board, then our modelling framework can be extended to include a model for the immigrant process.

Given that the immigrant $t_{i1}$ seeds cluster $i$, let $\mathbf t_I = \left(t_{11}, \dots, t_{S1}\right)$ denote the set of immigrant arrivals in $\mathbf y_{\tr}$, where immigrants are observed on the interval $\left( 0, a_0 \right]$. 
We then assume that the immigrant arrival $t_{i1}$ is generated by a Poisson process with intensity 
$$
\gamma_0 \left( t \right) = \lambda_0 \,  \boldsymbol \alpha_I' \boldsymbol s_{\boldsymbol \omega}\left( t \right),
$$
where $\lambda_0$ the intensity of immigrant arrivals over a full period and $\boldsymbol  \alpha_{I} ' \boldsymbol s_{\boldsymbol \omega}\left( t \right) = \alpha_I \left( t \right)$ is the absolute-time component of the immigrant intensity function, which is a periodic function with sinusoidal basis of the form defined in equation (\ref{eq:activity_function}).
The likelihood function for this model is 
$$
p \left( \mathbf tII \mid \lambda_0, \boldsymbol \alpha_I, \boldsymbol \omega \right) = \left[ \prod_{i = 1}^S \gamma_0 \left( t_{i1} \right)\right] \exp \left( - \Gamma_0 \left( a_0 \right)\right)
$$
where 
$$
\begin{aligned}
\Gamma_0 \left( a_0 \right) 
&= \int_{0}^{a_0} \gamma_0 (u) du, \\
&= \lambda_0 \int_{0}^{a_0} \boldsymbol \alpha' \boldsymbol s_{\boldsymbol \omega}\left( u \right) du, \\
&= \lambda_0 \, \boldsymbol \alpha' \boldsymbol S_{\boldsymbol \omega}\left( a_0 \right)
\end{aligned}
$$
for $S_0 \left( a_0 \right) = a_0$ and 
$$
S_{2k - 1} \left( a_0 \right) = \frac{1 - \cos \left( a_0 \omega_k \right)}{\omega_k} , \:
S_{2k} \left( a_0 \right) = \frac{\sin \left( a_0 \omega_k \right)}{\omega_k}. 
$$

\begin{figure}[ht]
    \centering
    \includegraphics[width=0.9\textwidth]{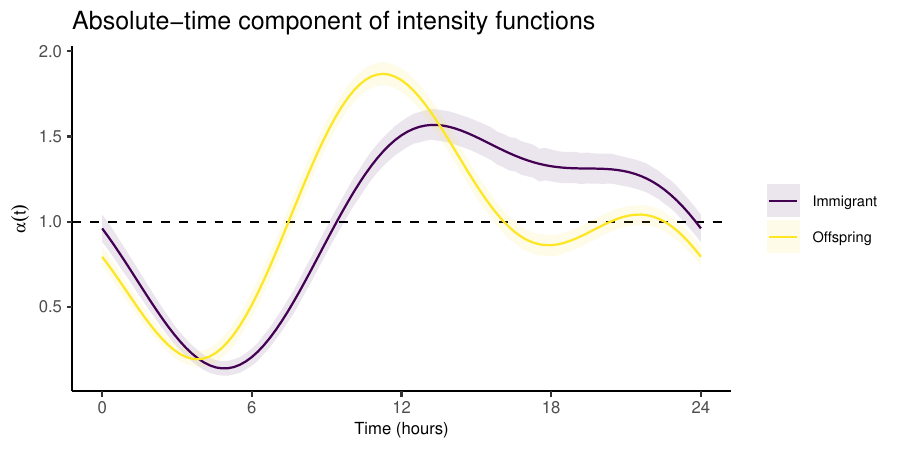}
    \caption{A comparison of the absolute-time component of the immigrant intensity function with the absolute-time component of the offspring intensity function inferred by $\mathcal{M}_4$. We see that these functions exhibit slightly different behaviour. This analysis implies that the intensity with which users post to the \texttt{r/ireland} subreddit peaks at approximately 14:00 but stays elevated until 23:00, while the intensity at which users make comments on existing posts peaks strongly at around 12:00.}
    \label{fig:immigrant_activity_function}
\end{figure}

Coding this model as a Stan program and fitting it to data is relatively straightforward, given priors for $\boldsymbol \alpha_I$ and $\lambda_0$ and hyper-parameters $\boldsymbol{\omega}$. In Figure \ref{fig:immigrant_activity_function} we present the immigrant function $\alpha_I \left( t \right)$ alongside the offspring function $\alpha \left( t \right)$ in $\mathcal{M}_4$. Here, we have specified $p \left( \alpha_I \right) = \mathcal N \left( 0, \sigma_\alpha^2 \right)$, $p \left( \log \lambda_0 \right) \propto 1$, and $\boldsymbol \omega = 2 \pi \boldsymbol f$ where $\boldsymbol f = \frac{1}{24} \left( 1, 2 \right)^\prime$. 

This sort of analysis can provide additional insights. For example, we see that the intensity function for posts is relatively flatter from 12:00 to 24:00 than that for comments, which has a clear peak around 12:00. That said, a detailed analysis of these patterns lies outside the scope of this study and is left for future work.

\end{appendix}

\end{document}